\documentclass[lettersize,journal]{IEEEtran}
\IEEEoverridecommandlockouts
\usepackage{graphicx, amsmath, amssymb, subcaption, url, cite, array, amsthm}
\usepackage[ruled, linesnumbered]{algorithm2e} 
\usepackage{algorithm2e, setspace}
\usepackage[font={small}]{caption}
\usepackage[hidelinks]{hyperref}
\graphicspath{ {./figures/} }
\DeclareMathOperator*{\argmax}{\arg\!\max}
\DeclareMathOperator*{\argmin}{\arg\!\min}

\begin{document}
\title{When Virtual Reality Meets Rate Splitting Multiple Access: A Joint Communication and Computation Approach}
\author{Nguyen Quang Hieu, Diep N. Nguyen, Dinh Thai Hoang, and Eryk Dutkiewicz}
\maketitle
\begin{abstract}
Rate Splitting Multiple Access (RSMA) has emerged as an effective interference management scheme for applications that require high data rates. Although RSMA has shown advantages in rate enhancement and spectral efficiency, it has yet not to be ready for latency-sensitive applications such as virtual reality streaming, which is an essential building block of future 6G networks. Unlike conventional High-Definition streaming applications, streaming virtual reality applications requires not only stringent latency requirements but also the computation capability of the transmitter to quickly respond to dynamic users’ demands. Thus, conventional RSMA approaches usually fail to address the challenges caused by computational demands at the transmitter, let alone the dynamic nature of the virtual reality streaming applications.
To overcome the aforementioned challenges, we first formulate the virtual reality streaming problem assisted by RSMA as a joint communication and computation optimization problem. A novel multicast approach is then proposed to cluster users into different groups based on a Field-of-View metric and transmit multicast streams in a hierarchical manner.
After that, we propose a deep reinforcement learning approach to obtain the solution for the optimization problem. Extensive simulations show that our framework can achieve the millisecond-latency requirement, which is much lower than other baseline schemes.

\end{abstract}

\section{Introduction}
\label{sec:introduction}
Recent years have witnessed the massive development of virtual reality, augmented reality applications as a part of the forthcoming 6G. With the significant increase in demands for such applications, the virtual reality sector is estimated to reach 18.6 billion dollars in revenue by 2026~\cite{vrstatistic2022}. Virtual reality applications require not only high data quality rendering ability of content providers but also stringent latency requirements with wireless connections~\cite{dang2020}. For applications such as 360-degree video streaming, the latency requirement is usually from a few to a hundred milliseconds, i.e., order of magnitude $10^{-3}$ second, to guarantee high Quality-of-Experience (QoE) for users~\cite{jiang2021}. Unlike conventional High-Definition (HD) video streaming, 360-degree video streaming also requires that the service provider can quickly respond to the changes in the Field-of-Views (FoVs) from the users. Moreover, the users are usually interested in a certain FoV, i.e., a fraction of the entire 360-degree frame, instead of roaming the whole 360-degree video frame~\cite{mangiante2017, perfecto2020}.
For this, the 360-degree video needs to be pre-rendered at the transmitter, e.g., at a Base Station (BS). This FoV pre-rendering process at the transmitter can not only guarantee quick responses to the changes of users' demands but also achieve spectral efficiency  through reductions of 360-degree video frames being transmitted~\cite{mangiante2017, perfecto2020, wei2021}. This FoV pre-rendering process is, however, a computational-expensive task at the transmitter~\cite{jiang2021}. 
Therefore, pre-rendering and transmitting 360-degree videos over wireless environments are critical in designing highly effective and reliable virtual reality streaming applications, especially when it comes to the diversity and heterogeneity of user equipments and user demands in future 6G networks~\cite{dang2020}.

The heterogeneity of the virtual reality equipment, e.g., head-mounted displays (HMDs), also poses a challenging problem in interference management for a large number of users with diverse interference levels. Considering the latency requirements in virtual reality streaming applications, most of the existing works utilize conventional multiple access schemes to address the interference among users~\cite{perfecto2020, wei2021 }. However, these works are either limited by the scheduling complexity due to time-slotted based multiple access~\cite{perfecto2020} or impractical by relying on the assumption of weak level of interference~\cite{wei2021}. 
To overcome the limitations due to high interference levels and high reliability requirements, Rate Splitting Multiple Access (RSMA) has been emerging as an efficient interference management scheme that partially decodes interference and partially treats interference as noise, and thus resulting in spectral and energy efficiency, reliability, and Quality-of-Service (QoS) enhancement~\cite{mao2018, dizdar2020, joudeh2017multicast}.
Research works in the literature have shown advantages of RSMA over conventional multiple access schemes, e.g., Spatial Division Multiple Access (SDMA) or Non-Orthogonal Multiple Access (NOMA) in terms of data rate, robustness, and energy efficiency~\cite{dai2016, mao2019}. 
Conventionally, these approaches only focus on the communication aspect of the transmitter, i.e., maximizing the sum-rate or min-rate fairness, under perfect/imperfect channel state information.
However, as discussed above, maximizing only the data rate does not guarantee the low latency requirement in such virtual reality applications since the latency is also caused by the computation processes, e.g., FoV pre-rendering, at the transmitter.
Such conventional approaches for RSMA thus cannot fully exploit the potential correlation between communication and computation, as well as the interactions between the transmitter and the users.
Therefore, the potential convergence of RSMA and virtual reality as a building block of the future 6G remains as an open research direction. 

To address the aforementioned problems, we propose a novel interference and computing resource management framework for virtual reality streaming applications assisted by RSMA.
Our main contributions can be summarized as follows.

\begin{itemize}
\item We propose a novel framework for low latency applications such as 360-degree video streaming with highly effective interference management assisted by RSMA. The latency requirement, interference management, and computing resource management are considered as a joint communication and computation optimization problem. 
The objective of the optimization problem is to minimize the system latency, subject to the resource constraints and dynamic of the system, i.e., channel state information, computing resource, and users' demands.
\item We then introduce innovative approaches to further improve the system resource usage as well as better manage interference among users. Specifically, we first propose a user clustering scheme that divides users into different groups based on  their overlapped FoVs. As such, we can significantly conserve the system resources since the transmitter only needs to transmit fractions of 360-degree video frames based on the groups' FoVs, rather than transmitting the entire 360-degree video frames to all the users. Second, we utilize the advantages of a Hierarchical RSMA (HRSMA) approach to transmit multicast streams to the groups of users. As such, users in different groups can be served based on their interests while the intra-interference among groups is efficiently managed.
\item To deal with the high-complexity as well as the dynamic and uncertainty of the considered optimization problem, we develop a highly-effective learning algorithm that is based on recent advances of machine learning techniques. For this, we collect a set of multimodal data regarding the communication and computation aspects of the transmitter. The multimodal dataset contains of (i) real-word dataset about users' behaviours, i.e., users' FoVs, watching different 360-degree videos, (ii) CPU usage of the transmitter running the FoV pre-rendering process, and (iii) imperfect channel state information at the transmitter.
A deep reinforcement learning (DRL) approach based on the Proximal Policy Optimization (PPO) algorithm is then developed to address the high complexity as well as effectively deal with the dynamic and uncertainty of the considered optimization problem. 
\item We carry out extensive simulation results evaluate the efficiency of our proposed framework compared to other conventional approaches, e.g., SDMA. In particular, our proposed framework can achieve latency in order of magnitude $10^{-3}$ second, given the uncertainty and heterogeneity of the considered environment. Moreover, we discuss and analyze various scenarios with real-world datasets together with real experiments to provide insightful and practical designs for virtual reality streaming applications assisted by RSMA.
\end{itemize} 

The rest of our paper is organized as follows. Section~\ref{sec:realted-works} discusses the related works and Section~\ref{sec:system-model} presents our proposed system model. Details of our approach based on deep reinforcement learning are described in Section~\ref{sec:drl-based-approach}. After that, our proposed pre-processing process and algorithm are described in Section~\ref{sec:training-drl}. Finally, Section~\ref{sec:performance-evaluation} discusses performance evaluation and Section~\ref{sec:conlcusion} concludes the paper.

\section{Related Works}
\label{sec:realted-works}
Motivated by the potential convergence of low-latency virtual reality applications and RSMA as a building block of the future 6G networks, we categorize related works into two main themes that are (i) virtual reality streaming supported by conventional multiple access schemes and (ii) integration of RSMA and virtual reality streaming applications.

\subsection{Virtual Reality Streaming with Multiple Access} 
In 360-degree video streaming, the video frames are usually divided into different tiles in which each tile is a rectangle containing a fraction of the entire 360-degree frame. A set of tiles then can be transmitted to the users' HMDs to display as users' FoVs. Considering that the FoVs of users usually have overlapped regions when the users experience the same video content, clustering users into groups before transmitting the tiles shows benefits in utilizing bandwidth and reducing latency~\cite{perfecto2020, wei2021}.
In~\cite{perfecto2020}, the authors propose a clustering framework based on users' locations for virtual reality applications. For this, the full knowledge of the seating area layout and the spatial locations of the small base stations as well as the users are assumed. A deep recurrent neural network model is then proposed to predict the FoVs of users and cluster users into different groups based on their FoVs and their locations. Simulation results show that clustering users based on FoV metric can reduce latency in a multicast scenario in which the FoVs being transmitted to groups can be reused. However, the use of time slotted multiple access scheme poses a complex scheduling problem with respect to the number of users in the network. It is shown in~\cite{perfecto2020} that as the number of users increases, e.g., from 3 to 5 users, the successful delivery ratio of the 360-degree video reduces significantly from above $70\%$ to less than $50\%$, respectively. To address the problem of time scheduling, the authors in~\cite{wei2021} propose a joint communication and computation framework for virtual reality video streaming applications based on SDMA. By utilizing multiple antennas at the transmitter,  the proposed approach shows advantages in enhancing QoE. However, the authors only consider an unicast scenario where the overlapped FoVs among users are not fully exploited. For example, when the number of users increases from 6 to 10 users, the QoE value decreases significantly from $60\%$ to $40\%$, respectively. 

It can be observed that most of current works on FoV-based video streaming either fail to utilize the reusable characteristics of the shared FoVs between users or cannot meet the stringent delivery and latency requirements due to limitations of time-slotted based multiple access. 
This thus calls for an innovative method to efficiently manage interference among users for virtual reality applications such as 360-degree video streaming. 
RSMA, with its advantages of partially decoding and partially cancelling interference, has been emerging as a potential approach to address the existing problems due to the following reasons. First, by utilizing the Successive Interference Cancellation (SIC) scheme at the receivers and rate splitting strategy at the transmitter, RSMA can achieve better data rate performance, resulting in higher QoE. Second, rate splitting strategy can be potentially utilized by transmitting different streams of data to the users based on their demands, e.g., users' FoVs. For this, advanced rate splitting strategies, e.g., multigroup multicast RSMA, can be considered as an effective solution for FoV-based video streaming.

\subsection{Integration of RSMA and Virtual Reality Streaming}
Similar to conventional multiple access approaches, e.g., SDMA and NOMA, the benefits of dividing users into different groups to deal with intra-interference (among groups) and inter-interference (among users within a group) attract vast attentions in RSMA research. 
In~\cite{joudeh2017multicast}, the authors introduce a novel multigroup multicast RSMA framework, in which each message is split into two parts that are a degraded part and a designated part. These parts are then encoded into corresponding degraded streams and designated streams. 
The proposed strategy can tackle the rate saturation problem in an overloaded scenario due to inter-group interference.
Simulation results show that the rate splitting strategy offers more flexibility through the degraded stream, thus marginally improving the rate performance. 
To further gain benefits from multigroup RSMA, the authors in~\cite{dai2016} propose a Hierarchical Rate Splitting Multiple Access (HRSMA) approach in which the messages are split and encoded into three streams that are (i) super common stream, (ii) group common stream, and (iii) private stream. By utilizing the super common streams and group common streams, the HRSMA approach can mitigate inter-group interference caused by overlapping channels, and intra-group interference caused by imperfect channel state information, respectively. Simulation results show that the HRSMA scheme outperforms other baselines in terms of sum-rate under various imperfect channel information scenarios. Other follow-up works have been proposed to utilize multigroup RSMA for various applications, e.g., multicell systems~\cite{tervo2018}, and satellite communications~\cite{yin2021, caus2018, zhu2018}. However, all the aforementioned works only focus on enhancing the rate performance of the RSMA networks through power allocation and precoder design approaches. 

Despite the fact that the user clustering methods can be either stationary~\cite{tervo2018, yin2021, caus2018} or dynamic~\cite{pereira2022}, the groups of users are usually formed based on the locations of users or the strength of channels. 
However, clustering users based on their locations or wireless channels might not be efficient in the context of 360-degree video streaming~\cite{vrinterference2022}. For example, clustering users based on FoVs can be beneficial in latency and spectral efficiency because the transmitter can only transmit a set of tiles of the 360-degree video frames~\cite{perfecto2020, mangiante2017}.
Moreover, allocating computing resources to users in RSMA is also another big challenge that has not been well investigated.
The computing resources of the system are usually dynamic due to the underlying processes run at the service provider~\cite{jiang2021}.
Due to the strict requirements of communication and computing constraints, finding optimization result for the problem is challenging under the uncertainty of the RSMA networks and the dynamic nature of virtual reality environment.

From all the above works and others in the literature, to the best of our knowledge, there is no work that can effectively address the challenges of streaming virtual reality applications taking account both interference management and computing resource constraints, under the dynamic and uncertainty of the virtual reality environment. Therefore, in this work, we propose a novel framework to address all the aforementioned problems. This framework is expected to not only effectively solve the joint communication and computation resource allocation problems for RSMA-assisted virtual reality streaming applications, but also extend to a wide range of applications in future 6G networks.

\section{System Model}
\label{sec:system-model}
\begin{figure}[t]
\centering
\includegraphics[width=1.0\linewidth]{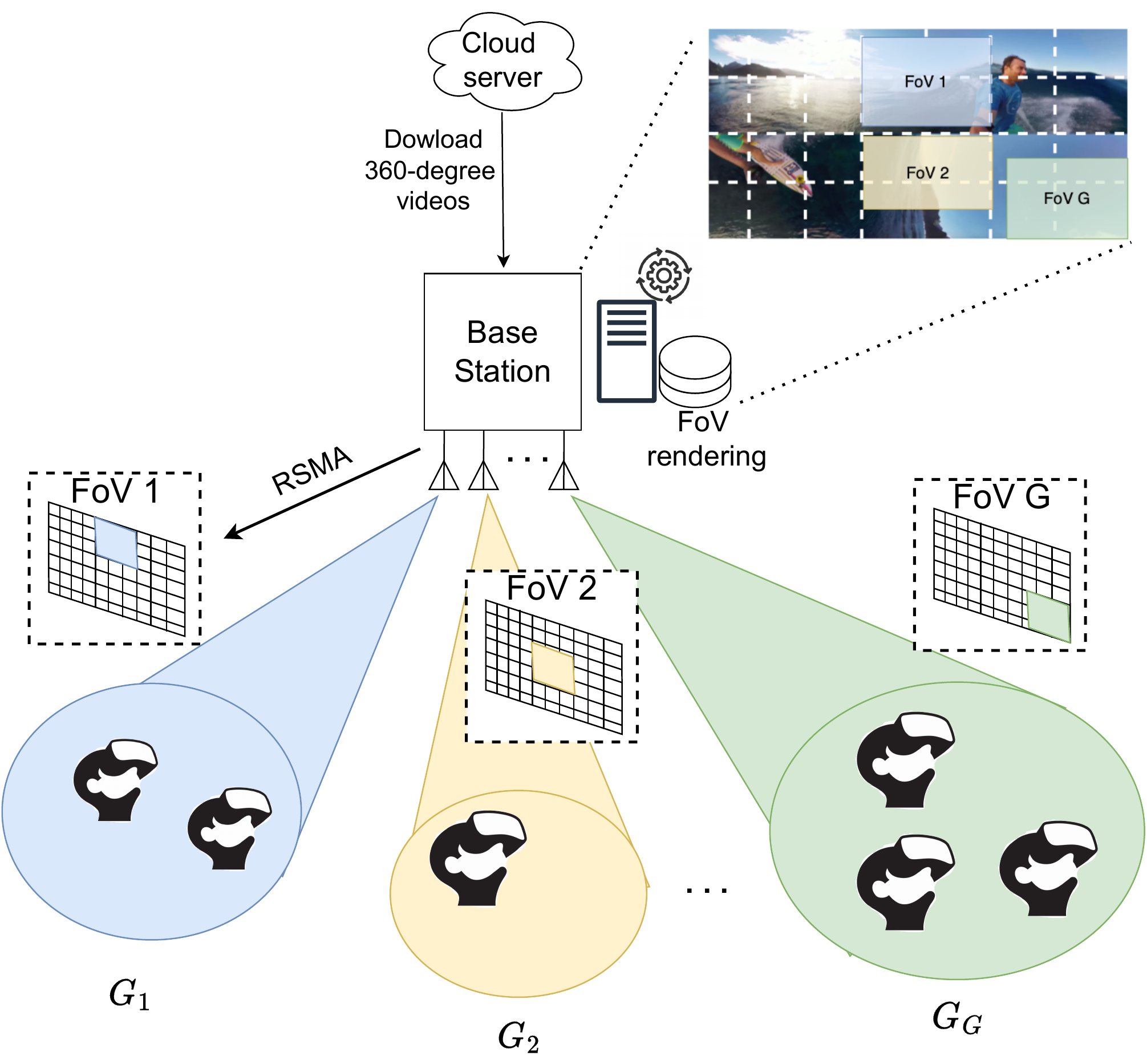}
\caption{An illustration of our system model. The $M$-antenna BS serves $K$ users belonging to $G$ groups in a 360-degree video streaming session. The BS has computing resources to pre-render FoVs for $G$ groups of users.}
\label{fig:system-model}
\end{figure}

Our proposed system model is illustrated in Fig.~\ref{fig:system-model} and can be described as follows. The system consists of $K$ users experiencing a 360-degree video streaming service provided by the $M$-antenna Base Station (BS) ($M \geq K$). Since multiple users are usually overlapped by FoVs, they can be clustered into $G$ groups based on their recent interests~\cite{perfecto2020}. Let $\mathcal{G} = \{G_1, G_2, \ldots, G_G\}$ denote the set of groups and $\mathcal{K} = \{1, \ldots, K\}$ denote the set of users. Once the user groups are formed, the BS splits and transmits messages (in the bit-level) to $K$ users and $G$ groups as follows.
The message $W_k$ intended for user $k$ belonging to group $G_g$ is split into three parts that are a super common message, i.e., $W_k^c$, a private message, i.e., $W_k^p$, and a group common message $W_k^{G_g}$. The first part $W_k^c$ is encoded into the super common stream $z_c$. The super common stream $z_c$ contains common messages of $K$ users, i.e., $W_k^c, \forall k \in \mathcal{K}$. The private message $W_k^p$ is encoded into a private stream $z_k$. The group common message $W_k^{G_g}$ intended for group $G_g$ is encoded into a stream $z_{G_g}$. The transmit stream of the BS is $z = (z_c, z_{G_1}, z_{G_2}, \ldots, z_{G_G}, z_1, z_2, \ldots, z_K)$. By utilizing multiple antennas at the BS, $K$ streams $z$ are simultaneously transmitted to $K$ users. 
Let $P_0$ denote the transmission power at the BS, $\alpha_c$, $\alpha_{G_g}$, and $\alpha_k$ denote the power coefficients allocated to stream $z_c$, stream $z_{G_g}$, and stream $z_k$, respectively. The transmit signal at the BS is~\cite{mao2018}:

\begin{equation}
\mathbf{x} = \sqrt{P_0}\left(\sqrt{\alpha_c} z_c + \sum_{G_g \in \mathcal{G}} \sqrt{\alpha_{G_g}} z_{G_g} + \sum_{k \in \mathcal{K}} \sqrt{\alpha_k} z_k \right).
\end{equation}
The received signal at user $k$ is:

\begin{equation}
y_k = \mathbf{h}_k^{H} \mathbf{x} + w_k, \forall k \in \mathcal{K},
\end{equation}
where $\mathbf{h}_k \in \mathbb{C}^{M \times 1}$ is the channel between the user $k$ and the BS, and $w_k$ is the Additive White Gaussian Noise (AWGN) noise, i.e., $w_k \sim \mathcal{CN}(0, \sigma_k^2)$.

Each user then decodes three streams $z_c$, $z_{G_g}$, and $z_k$. The super common stream is first decoded by treating interference from other streams as noise. The SINR of decoding the super common stream at user $k$ is:

\begin{equation}
\gamma_k^c = \frac{|\mathbf{h}_k|^2 P_0 \alpha_c}{|\mathbf{h}_k|^2 \sum_{G_g \in \mathcal{G}} P_0 \alpha_{G_g} + |\mathbf{h}_k|^2 \sum_{k \in \mathcal{K}} P_0 \alpha_k + 1}.
\end{equation}

Once the super common stream is decoded, it is subtracted from the received signal $y_k$ by using SIC. After that, the user $k$ decodes its group common stream $z_{G_g}$ by treating interference from other group common streams and other private streams as noise. The SINR of decoding the group common stream at user $k$ is:

\begin{equation}
\gamma_k^{G_g} =  \frac{|\mathbf{h}_k|^2 P_0 \alpha_{G_g}}{|\mathbf{h}_k|^2 \sum_{G_{g'} \in \{\mathcal{G} \setminus G_g\}} P_0 \alpha_{G_{g'}} + |\mathbf{h}_k|^2 \sum_{k \in \mathcal{K}} P_0 \alpha_k + 1}.
\end{equation}

After decoding the corresponding group common stream, user $k$ decodes its private stream $z_k$. The SINR of decoding the private stream at user $k$ is:

\begin{equation}
\gamma_k^p = \frac{|\mathbf{h}_k|^2 P_0 \alpha_k}{|\mathbf{h}_k|^2 \sum_{G_{g'} \in \{\mathcal{G} \setminus G_g\}} P_0 \alpha_{G_{g'}} \!+\! |\mathbf{h}_k|^2 \sum_{k' \in \{\mathcal{K} \setminus k\}} P_0 \alpha_{k'}\! + \!1}.
\end{equation}

Given the above SINRs, the achievable data rate for each user $k$ can be calculated as a sum of (i) the rate of decoding the private stream, denoted as $R_k$, (ii) the rate of decoding a portion of the group common stream, denoted as $C_k^{G_g}$, and (iii) the rate of decoding a portion the super common stream, denoted as $C_k^{c}$. Thus, the achievable rate of the user $k$ in group $G_g$ is:

\begin{equation}
R_k^{G_g} = R_k^p + C_k^{G_g} + C_k^c.
\label{eq:user-rate}
\end{equation}

Because each user decodes a portion of the group common stream and a portion of the super common stream, we can obtain the rate constraints for the super common stream and group common streams as follows:
\begin{equation}
\sum_{k \in \mathcal{K}} C_k^c = R_c,
\end{equation}

\begin{equation}
\sum_{k \in \mathcal{K}} C_k^{G_g} = R_{G_g}, \forall  G_g \in \mathcal{G},
\end{equation}
where $R_{G_g}$ and $R_c$ are achievable rates of user $k$ for decoding the group common stream and super common stream, respectively. $R_k$, $R_{G_g}$, and $R_c$ can be calculated as follows.

\begin{equation}
R_k = \log_2(1+\gamma_k^p),
\end{equation}

\begin{equation}
R_{G_g} = \min_{k \in \mathcal{K}, G_g \in \mathcal{G}}\Big\{ \log_2(1+\gamma_k^{G_g}) \Big\} ,
\end{equation}

\begin{equation}
R_c = \min_{k \in \mathcal{K}} \Big\{\log_2(1+\gamma_k^{c}) \Big\}.
\end{equation}

\begin{figure}[t]
    \centering
    \begin{subfigure}[t]{0.49\linewidth}
        \includegraphics[width=1.0\linewidth]{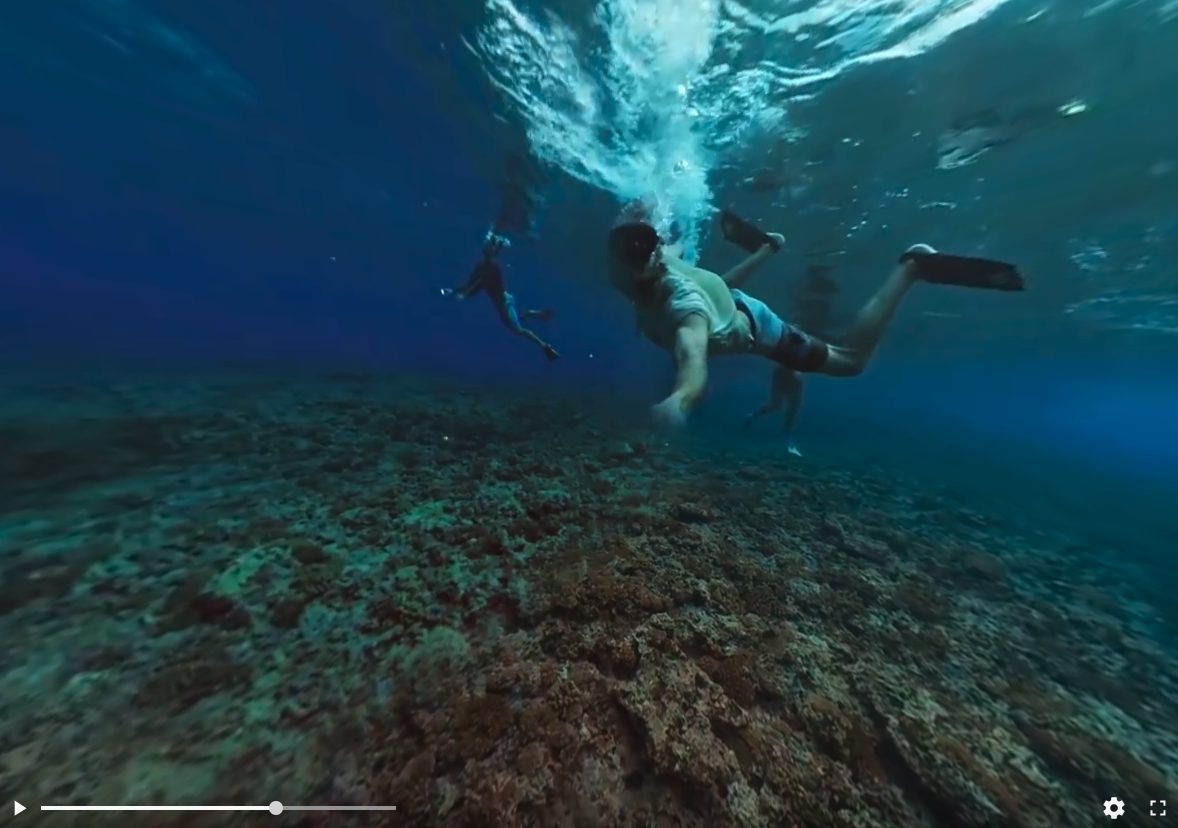}
        \caption{}
    \end{subfigure}%
    ~ 
    \begin{subfigure}[t]{0.49\linewidth}
       \includegraphics[width=1.0\linewidth]{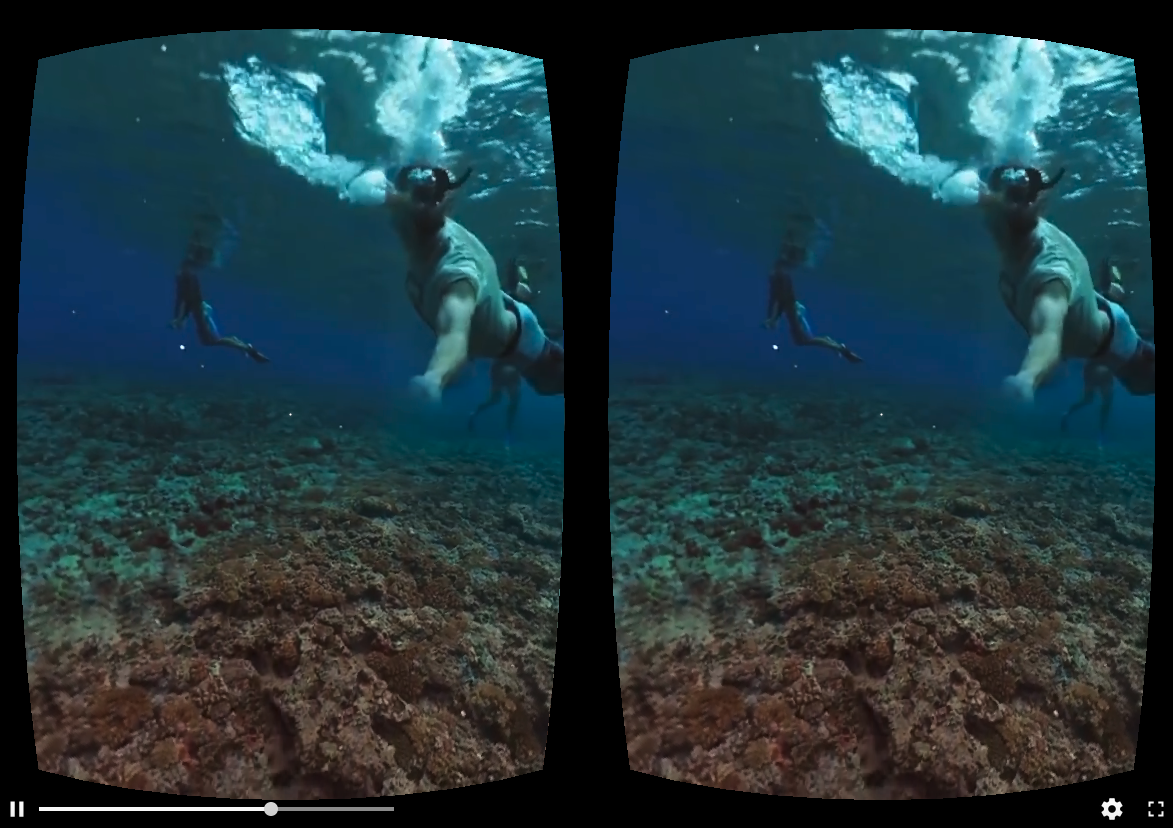}
       \caption{}
    \end{subfigure}
    \caption{FoV pre-rendering at the BS: (a) a selected video frame before rendering and (b) after rendering. The video is selected from the dataset~\cite{dharmasiri2021} and publicly available on Youtube\protect\footnotemark. Our local server for running the FoV pre-rendering process is a MacBook Air 2020 with 8GB memory and 2.3 GHz 8-core CPU.}
    \label{fig:fov-render}
\end{figure}

In streaming 360-degree videos, the latency is a major consideration for users' QoE. We thus measure the total latency of the system based on (i) the latency caused by video transmission and (ii) the latency caused by FoV pre-rendering process at the BS. 
In the computational perspective, rendering FoVs at the BS is the most heavy computing task~\cite{mangiante2017}, and thus we assume that the latency for other computing tasks, e.g., storing dataset and dividing messages for RSMA, is negligible.
In this case, the video transmission latency of user $k$ can be calculated by:

\begin{equation}
L_{v,k} = \frac{N_x}{B R_k^{G_g}},
\end{equation}
where $N_x$ (bits) is the size of the transmitted signal $\mathbf{x}$ and $B$ (Hz) is the bandwidth of the broadcast channel. To measure the FoV pre-rendering latency at the BS, we set up a real experiment as follows. We first collect the dataset that contains 360-degree videos and users' behaviours data experiencing these videos from~\cite{dharmasiri2021}. The FoV for each group $G_g \in \mathcal{G}$ is rendered by using the open-source software Vue-VR~\cite{ibrahim2022}. Fig.~\ref{fig:fov-render} illustrates a selected video frame before and after rendering with Vue-VR.

    \footnotetext{https://www.youtube.com/watch?v=MKWWhf8RAV8}

\begin{figure}[t]
\centering
\includegraphics[width=0.8\linewidth]{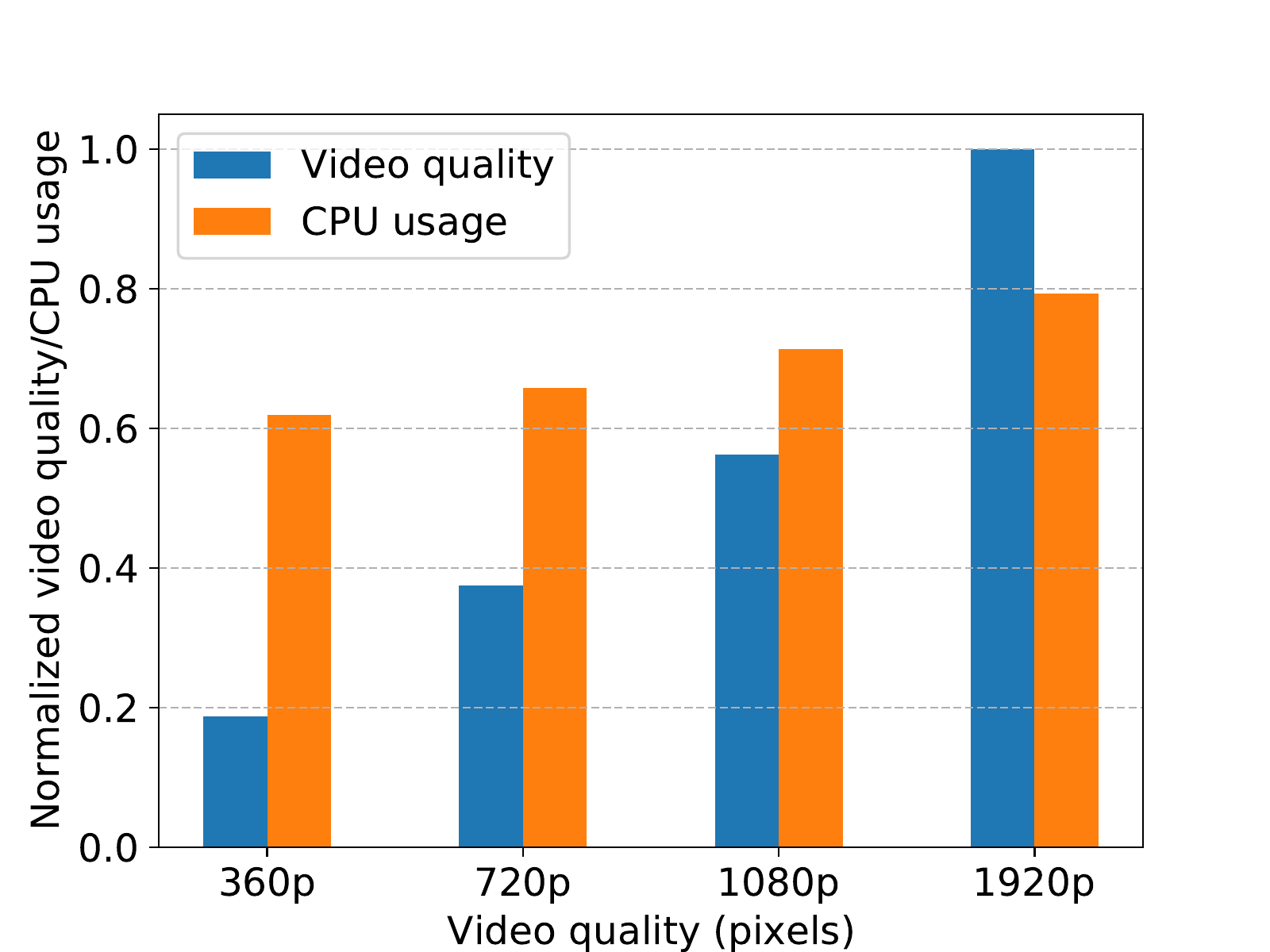}
\caption{Correlation between CPU usage and video quality of the FoV pre-rendering process at our local sever. The CPU information of the local server is measured every $t=2$ seconds. 
We use the same 360-degree video as in Fig.~\ref{fig:fov-render} with four different qualities for the experiment. The number of collected samples corresponding to each video quality is 11,000, which is equivalent to 6 hours.
}
\label{fig:cpu-latency}
\end{figure}

We measure the CPU usage of the local server running FoV pre-rendering process with a selected 360-degree video under different qualities. The experimental results are illustrated in Fig.~\ref{fig:cpu-latency}. It can be observed from this experiment that when the video quality increases, the BS spends more CPU resources for rendering the FoVs for the groups.
From the obtained CPU usage at the BS, we can derive the FoV pre-rendering latency for user $k$ in group $G_g$ as follows. Let $f_{g} \leq 1$ denote the proportion of CPU cycles that the BS spends on rendering FoV for group $G_g$ and $F_g \leq 100\%$ denote the current capacity that the BS can spend on this FoV rendering task. The values for $F_g$ are obtained from the experiment as shown in Fig.~\ref{fig:cpu-latency}. The FoV pre-rendering latency for user $k$, denoted as $L_{r,k}$ in group $G_g$ can be calculated by:

\begin{equation}
L_{r,k} =  \frac{\mathbf{1}_k(\pi_G)}{f_{g} F_g \eta},
\end{equation}
where $\eta$ (Hz) is the CPU capacity, i.e., total number of computing cycles, of the BS, and $\mathbf{1}_k(\pi_G)$ is the indicator function, given a current clustering policy $\pi_G$. In particular, $\mathbf{1}_k(\pi_G) = 1$ if the user $k$ is in the group $G_g$. Otherwise $\mathbf{1}_k(\pi_G) = 0$.
Once we obtain the values of the video transmission latency and FoV pre-rendering latency, the total latency of the user $k$ can be calculated as follows:

\begin{equation}
L_{s,k} = L_{v,k} + L_{r,k}.
\end{equation}

To guarantee the fairness among all users, the goal of the BS is to minimize the latency of user that has the longest delay, denoted as ``max-latency" hereafter.
Let $\pi_G$ denote the clustering policy of the BS and $\Psi_G$ denote the set of all clustering policies, the optimization problem can be formulated as follows:

\begin{subequations}
\label{eq:min-latency}
\begin{align}
\min_{\boldsymbol{\alpha}, \mathbf{C}, \mathbf{f}} \quad & \max_{k \in \mathcal{K}} L_{s,k} \\
\textrm{s.t.} \quad & \alpha_c + \sum_{G_g \in \mathcal{G}} \alpha_{G_g} + \sum_{k \in \mathcal{K}} \alpha_k \leq 1,\\
\quad & \sum_{k \in \mathcal{K}} C_k^c = R_c,  \\
\quad & \sum_{k \in \mathcal{K}} C_k^{G_g} = R_{G_g}, \forall G_g \in \mathcal{G},  \\
\quad & \pi_G \in \Psi_G, \\
\quad & f_{g} \leq 1,
\end{align}
\end{subequations}
where $\boldsymbol{\alpha} = [\alpha_c, \alpha_{G_1}, \ldots, \alpha_{G_G}, \alpha_1, \ldots, \alpha_K]$ is the power allocation coefficient vector, $\mathbf{C} = [C_1^c, \ldots, C_K^c, C_1^{G_1}, \ldots, C_K^{G_1}, \ldots, C_1^{G_G}, \ldots, C_K^{G_G}]$ is the rate allocation vector, and $\mathbf{f} = [f_1, \ldots, f_G]$ is the computing resource allocation vector.
In the above optimization problem, (\ref{eq:min-latency}b) is the power constraint that ensures the power used for communication does not exceed the BS's power budget. (\ref{eq:min-latency}c) and (\ref{eq:min-latency}d) are the super common rate and group common rate constraints, respectively, to guarantee that the rates allocated for users and groups are decodable at the receivers.
(\ref{eq:min-latency}e) is the clustering policy constraint.
Finally, (\ref{eq:min-latency}f) is the computing resource constraint to ensure that the proportion of CPU cycles used for FoV pre-rendering process for each group does not exceed the CPU capacity.

It can be observed that the problem in (\ref{eq:min-latency}) is a form of the multigroup multicast power control (MMPC) problem in~\cite{karipidis2008} if we remove the component $L_{r,k}$ in the optimization objective $L_{s,k}$. As such, solving the problem (\ref{eq:min-latency}) can be reduced to the MMPC problem, which is known to be NP-hard~\cite{karipidis2008}. 
In other words, even ignoring the dynamics and uncertainties of the  the wireless channels and/or the users' interests (measured in the dynamic clustering policy), it is intractable to solve (\ref{eq:min-latency}) for its optimal solution. Note that the FoV pre-rendering process at the BS also causes more uncertainties for the system. 

To find the solution for (\ref{eq:min-latency}), conventional optimization approaches for RSMA in the literature, e.g.,~\cite{mao2018, dai2016, joudeh2017multicast, zhu2018} and~\cite{yin2021}, cannot be directly applied due to the following reasons. First, such approaches can only apply for the scenarios in which the distribution of the channel state information is known in advance. 
Second, unlike conventional sum-rate or max-min fairness optimization problems in~\cite{mao2018, dai2016, joudeh2017multicast, zhu2018}, and~\cite{yin2021}, our optimization problem in (\ref{eq:min-latency}) is even more challenging because we consider both unknown distribution of the channel state information and nonstationary processes caused by the CPU usage of the BS and the users' demands.
Therefore, this requires a novel solution to deal with dynamic nature and unknown distributions of the system to find the optimal solutions.
In the following section, we propose a novel learning approach based on deep reinforcement learning (DRL) that can not only deal with the uncertainty of the parameters but also the heterogeneity of the underlying processes. In particular, we first transform the optimization problem in (\ref{eq:min-latency}) into a policy optimization problem in a DRL setting. We then define the fundamental parameters to design an effective DRL approach.

\section{DRL-based 360-degree Video Streaming with RSMA}
\label{sec:drl-based-approach}
Before explaining our proposed framework in detail, we first introduce the fundamental background of DRL.

\subsection{Deep Reinforcement Learning}
Deep Reinforcement Learning is a combination of deep neural networks and reinforcement learning (RL)~\cite{mnih2013}. 
In conventional RL settings, a learning agent aims to obtain an optimal policy $\Omega^*$ through interacting with a given dynamic environment in discrete time steps. At each time step $t$, the agent observes its current state $s_t$ in a state space $\mathcal{S}$. Based on the observed state $s_t$, the agent takes an action $a_t$ in an action space $\mathcal{A}$, then transits to the next state $s_{t+1}$, and receives a reward signal $r_t \in \mathbb{R}$. The reward signal $r_t$ is a feedback from the environment that indicates how good the agent is performing at the current state. As a result, the optimal policy of the agent can be obtained by maximizing the long-term average reward, i.e.,

\begin{equation}
\Omega^* = \argmax_{\Omega} \frac{1}{T}\sum_{t=0}^T r_t,
\label{eq:rl-max-policy}
\end{equation}
where $T$ is the time horizon. The policy $\Omega$ can be defined by a probability distribution over actions, i.e., $\Omega = \text{Pr} \{a_t|s_t\}$.

In conventional RL settings as described above, finding the optimal policy for (\ref{eq:rl-max-policy}) usually faces a policy search problem in which the convergence time of the RL agent depends on the search spaces $\mathcal{S}$ and $\mathcal{A}$. In large discrete spaces that contain thousands of state/action values, or spaces that are continuous, finding the optimal policy can be very time-consuming or infeasible~\cite{sutton2018}.
To overcome the policy search problem, deep neural networks have been integrated into RL as a nonlinear function approximator~\cite{mnih2013, silver2014, schulman2017, sutton2018}, which is known as DRL. DRL shows significant improvements over RL in terms of convergence time and better policy. In DRL, the policy can be illustrated as $\Omega = \text{Pr}\{a_t|s_t;\boldsymbol{\theta}\}$, where $\boldsymbol{\theta}$ is the parameter vector, i.e., weights and biases, of the deep neural networks. 
The training procedure and details of our proposed DRL-based framework are described as follows.

\subsection{DRL-based Optimization Formulation}
To convert the optimization problem (\ref{eq:min-latency}) into a problem that can be optimized by DRL, we define the state space, action space, reward signal of the DRL agent as follows. The DRL agent can be deployed at the BS to observe the surrounding environment and make corresponding actions.
The state space of the BS can be defined as follows:

\begin{equation}
\mathcal{S} = \Big\{ \hat{h}_k, F_g, \pi_G; \forall k \in \mathcal{K}, g \in \{1, 2, \ldots, G\}, \forall \pi_G \in \Psi_G \Big\},
\end{equation}
where $\hat{h}_k \in \mathbb{R}$ is the root mean square of the channel gain $\mathbf{h}_k$ between user $k$ and the BS, $F_g \in [0, 1]$ is maximum portion of CPU allowed to render FoV for group $G_g$, and $\pi_G$ is the current clustering policy (i.e., which user is currently belonging to which group).
Details of the components in the state space are as follows.
The clustering policy $\pi_G$ is a $G \times K$ binary matrix that has $G$ entries in which each entry is equivalent to the state of users in the corresponding group. For example, given $G=3$ groups and $K=6$ users, we can derive the clustering policy $\pi_G$ as follows:

\begin{equation}
\pi_G = \begin{bmatrix}
0 & 1 & 0 & 0 & 0 & 1\\
0 & 0 & 0 & 1 & 0 & 0\\
1 & 0 & 1 & 0 & 1 & 0\\
\end{bmatrix},
\end{equation}
where the first entry, i.e., first row, illustrates that the second and sixth users are currently in the first group. Similarly, the fourth user is in the second group and the first, third, and fifth users are in the third group.
The channel gain between the user $k$ and the BS is defined by:

\begin{equation}
\mathbf{h}_k = \mathbf{\bar{h}}_k + \tilde{h}_k, 
\end{equation}
where $\mathbf{\bar{h}}_k \in \mathbb{C}^{M \times 1}$ is the constant channel gain and $\tilde{h}_k \in \mathcal{CN}(0, \sigma_k^2)$ is the feedback error from the user. 
The action space of the BS is defined as follows:

\begin{equation}
\mathcal{A} = \{\boldsymbol{\alpha}, \mathbf{C}, \mathbf{f}\},
\end{equation}
where $\boldsymbol{\alpha}$, $\mathbf{C}$, and $\mathbf{f}$ are the power allocation coefficient vector, rate allocation vector, and computing resource allocation vector as defined in Section~\ref{sec:system-model}.
At time step $t$, the BS observes the current state $s_t \in \mathcal{S}$, executes action $a_t \in \mathcal{A}$ based on the current policy $\Omega$, and obtains the reward signal $r_t(s_t, a_t)$ at the end of the time step.
It is noted that the BS's goal is minimizing the upper bound latency (max-latency) of the system, i.e., $\max_{k \in \mathcal{K}} L_{s,k}$, meaning that the upper bound latency for all users can be guaranteed. 
Since the latency is a positive value, we can transform this max-latency minimization problem in (\ref{eq:min-latency}) into maximizing the inverse of the max-latency.
Therefore, the reward signal of the BS is defined as the inverse of the max-latency, i.e., 

\begin{equation}
r_t(s_t, a_t) = \frac{1}{\max_{k \in \mathcal{K}}L_{s,k}}.
\label{eq:reward}
\end{equation}
By defining the reward signal as above, finding an optimal policy for (\ref{eq:rl-max-policy}) is equivalent to max-latency minimization problem in (\ref{eq:min-latency}).

From the BS's perspective, the above procedure can be described as follows. At the beginning of time step $t$, the BS observes the current state of the surrounding environment $s_t$ by (i) storing channel feedback $\mathbf{H} = [\mathbf{h}_1, \ldots, \mathbf{h}_K]$ from all users, (ii) estimating the amount of CPU used to render FoVs for groups, and (iii) updating the current clustering policy. Based on the observed state, the BS then performs action $a_t$ that consists of (i) power allocation coefficient vector, (ii) rate allocation vector, and (iii) computing resource allocation vector. Once the action $a_t$ is executed, the BS receives the reward signal $r_t$ at the end of the time step. Note that unlike other conventional  optimization approaches~\cite{mao2018, dai2016, mao2019}, our algorithm does not require an exact objective function or distributions of the parameters, which are usually hard to acquire in practice, to obtain the optimization results. Our algorithm only requires the reward signals which are in the form of feedback from users at the end of the decision time step. Similar to~\cite{wei2021} and~\cite{perfecto2020}, we assume that the feedback from users also contains the information about users' FoVs which are used for the user clustering purpose. Since the FoVs of the users can only be obtained at the end of the decision time step, the BS needs to make an action that takes into consideration the possible changes of users' FoVs in the new time step~\cite{perfecto2020}.

By defining the state space, action space, and reward signal as above, the optimization problem in (\ref{eq:min-latency}) can be transformed into finding the optimal policy $\Omega$ as follows:
\begin{subequations}
\begin{align}
\quad & \max_{\Omega}\mathbb{E}_{a_t, s_t} \Big[\frac{1}{T} \sum_{t=0}^T r_t\Big] \\
\textrm{s.t.}  \quad & a_t \in \mathcal{A}, \\
\quad & s_t \in \mathcal{S}, \\
\quad & a_t \sim \Omega(a_t, s_t;\boldsymbol{\theta}),\\
\quad & s_t \sim \mathcal{P}(s_{t+1}|s_t, a_t),
\end{align}
\label{eq:drl-optimization}
\end{subequations} 
where $\mathcal{P}(s_{t+1}|s_t, a_t)$ is the state transition probability which indicates the dynamic of the environment. Note that the state transition probability is unknown to the BS. The BS can only observe its surrounding environment and receive reward signal from the environment as described above.

Solving (\ref{eq:drl-optimization}) is challenging as the action space is a large multi-dimensional, continuous space. In addition, the state space consists of multimodal data with unknown distributions, i.e., CPU usage of the BS, imperfect channel state information, and users' behaviours.
To solve the problem in (\ref{eq:drl-optimization}), we develop a highly-effective DRL approach based on the Proximal Policy Optimization (PPO) algorithm~\cite{schulman2017}. To deal with large continuous action space, the PPO algorithm utilizes a policy clipping technique to guarantee a robust training convergence. In addition, the heterogeneity of input data can be addressed by using deep neural networks as nonlinear function approximators. 
The details of our approach are described in the next section.

\section{Training DRL Model}
\label{sec:training-drl}
\begin{figure*}[t]
\centering
\includegraphics[width=1.0\linewidth]{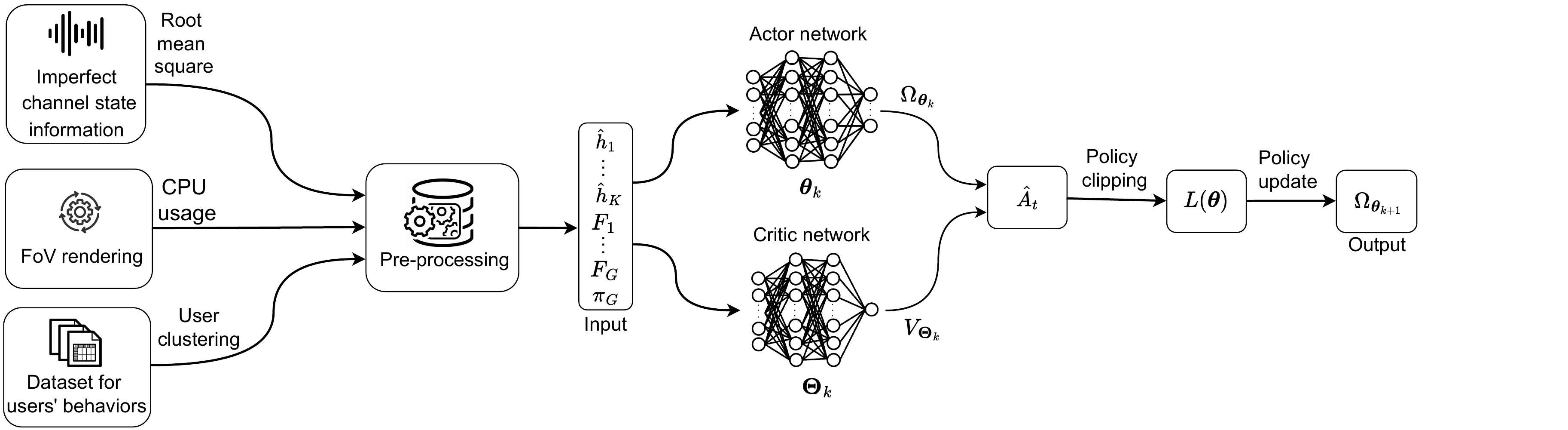}
\caption{Data pre-processing and training pipeline in our DRL-based approach.
The multimodal data from three different sources are collected from the experiments as shown in Fig.~\ref{fig:cpu-latency} and~\cite{dharmasiri2021}.
The data is then pre-processed to synchronize sampling rates and remove abnormal data points. Finally, the processed data is iteratively fed into the deep neural networks to update the parameters of those deep neural networks.
After the training process finishes, the trained model can be applied to the new i.i.d data in the testing process.
}
\label{fig:ppo-update}
\end{figure*}

\subsection{Data Processing}
Our proposed learning approach is illustrated in Fig.~\ref{fig:ppo-update}. In order to prepare for the training process with deep reinforcement learning, we first propose the data pre-processing process as follows. The state space of the BS is constructed from three different data sources. The first part of the state space of the BS consists of the root mean square values of the channel gains, i.e., $\hat{h}_k, \forall k \in \mathcal{K}$. The root mean square of the channel gain is widely adopted to train learning models in the literature~\cite{ye2017}. The channel gain $\mathbf{h}_k$ is a complex variable as defined in Section~\ref{sec:system-model}. To capture the CPU usage of the BS under the FoV pre-rendering process, we run the Vue-VR software~\cite{ibrahim2022} in a local server as described in Section~\ref{sec:system-model}. The CPU state of the local server is captured at a sampling rate that is equal to the length of a time step $t$. To divide users into different groups based on their FoVs, we utilize a dataset about users' behaviours from~\cite{dharmasiri2021}. With 17 categorized features, we apply the K-means cluster algorithm to obtain clustering results for $G$ groups of user. The details of the categorized features and clustering results are described in Section~\ref{subsec:simulation-settings}.

Once we collect all the components of the state space for the BS, we pre-process all the data before the data can be fed into the DRL model. The pre-processing is to remove abnormal data points, e.g., infinite channel gains or CPU state values that exceed $100\%$. After the data of the state space is processed, the data entries are iteratively fed into two streams of the DRL model as shown in Fig.~\ref{fig:ppo-update}. The details of the training process with DRL-based learning approach are described in the following section. 

\subsection{Proximal Policy Optimization}
\begin{algorithm}[t]
\caption{Proximal Policy Optimization (PPO)}
 \textbf{Input}: \\
 Initialize parameter vector $\boldsymbol{\theta}_{0}$ for actor network, \\ 
 Initialize parameter vector $\boldsymbol{\boldsymbol{\Theta}}_{0}$ for critic network,  \\
 \For{$k = 0, 1, 2, \ldots$}{
  Collect set of trajectories $\mathcal{B}_{k}=\left\{\upsilon_{t};\upsilon_t = (s_t, a_t, r_t)\right\}$ by running policy $\Omega_{\boldsymbol{\theta}_k}$ in the environment \\
  Compute cumulative reward $\hat{R}_{t} = \sum_{t=0}^{T} r_t$ \\
  Compute advantage function $\hat{A}_{t}$ as in (\ref{eq:advantage-function}) \\
  Update the policy by maximizing the objective (\ref{eq:ppo-loss}): 
   \begin{equation*}
\boldsymbol{\theta}_{k+1}=\argmax_{\boldsymbol{\theta}} \frac{1}{\left|\mathcal{B}_{k}\right| T} \sum_{\upsilon \in \mathcal{B}_{k}} \sum_{t=0}^{T} L(\boldsymbol{\theta}),
 \end{equation*} \\
  Fit value function by regression on mean-squared error: 
  \begin{equation*}
\boldsymbol{\Theta}_{k+1}=\argmin_{\boldsymbol{\Theta}} \frac{1}{\left|\mathcal{B}_{k}\right| T} \sum_{\upsilon \in \mathcal{B}_{k}} \sum_{t=0}^{T}\left(V_t\left(s_{t};\boldsymbol{\Theta}\right)-\hat{R}_{t}\right)^{2}
 \end{equation*}
 }
 \textbf{Outputs}: $\Omega_{\boldsymbol{\theta}}^* = \text{Pr}(a_t|s_t;\boldsymbol{\theta})$
\label{algo:ppo}
\end{algorithm}

Our proposed learning approach is developed based on Proximal Policy Optimization (PPO) algorithm~\cite{schulman2017}. There are two main reasons for using PPO in this work. First, our problem in (\ref{eq:drl-optimization}) requires the continuous action space so that the conventional action-value methods, e.g., Deep Q-Network (DQN)~\cite{mnih2015} or Deep Dueling Q-network (DDQN)~\cite{ wang2016}, cannot be directly used. Second, PPO can guarantee stable training results in high dimensional spaces, thanks to its clipping technique on the gradient step updates. Details of the training process are as follows.

PPO estimates the policy of the BS by using two deep neural networks, i.e., an actor network and a critic network. The actor network, denoted as $\Omega_{\boldsymbol{\theta}_k}$, is used for the output policy, i.e., power allocation, rate control, and computing resource allocation for the BS. The critic network, denoted as $V_{\boldsymbol{\Theta}_k}$, is used for evaluating how good or bad the current state of the BS is. The details of the training process for updating parameters of the two networks are shown in Algorithm 1 and can be described as follows.
The parameter vectors, i.e., weights and biases, of the actor network and critic network are initialized (lines 2 and 3 in Algorithm 1). Each policy update, indexed by $k$, is performed as follows. A set of trajectories, i.e., $(s_t, a_t, r_t)$, is collected by running the current policy $\Omega_{\boldsymbol{\theta}_k}$ in the environment for $T$ time steps. The cumulative reward is calculated by $\hat{R}_{t} = \sum_{t=0}^{T} r_t$. Based on the obtained outputs of the actor network and critic network, an advantage function, denoted as $\hat{A}_t$, can be calculated as follows:

\begin{equation}
\hat{A}_t(s_t, a_t; \boldsymbol{\theta}) = Q_t(s_t, a_t; \boldsymbol{\theta}) - V_t(s_t; \boldsymbol{\Theta}),
\label{eq:advantage-function}
\end{equation}
where $Q_t(s_t, a_t; \boldsymbol{\theta}) = \mathbb{E}_{a_t \in \Omega, s_t \in \mathcal{P}}\Big[\sum_{l=0}^{T - t}r_t(s_{t+l}, a_{t+l})\Big]$ is the action value function and $V_t(s_t; \boldsymbol{\Theta}) = \mathbb{E}_{s_t \in \mathcal{P}}\Big[\sum_{l=0}^{T-t} r_t(s_{t+l}, a_{t+l})\Big]$ is the state value function. The action value function and state value function of the actor network and critic network, respectively, can be calculated by the expectation, i.e., $\mathbb{E}(\cdot)$, over $l$ time steps from the current time step $t$ when the BS follows the current policy $\Omega_{\boldsymbol{\theta}_k}$. The advantage function $\hat{A}_t$ estimates whether the action taken, i.e., output of the actor network, is better than the policy's default behaviour.
Once the value of the advantage function is obtained, the training objective function can be calculated by:

\begin{equation}
L(\boldsymbol{\theta}) = \min\Big(\frac{\Omega_{\boldsymbol{\theta}_{k}}}{\Omega_{\boldsymbol{\theta}_{k-1}}} \hat{A}_t, u(\epsilon, \hat{A}_t)\Big),
\label{eq:ppo-loss}
\end{equation}
where $u(\epsilon, \hat{A}_t)$ is the clip function that limits significant gradient step updates which may degrade the performance of the BS. The clip function is defined as follows~\cite{schulman2017}:

\begin{equation}
u(\epsilon, \hat{A}_t) =
    \begin{cases}
      (1 + \epsilon) \hat{A}_t, & \text{if $\hat{A}_t \geq 0$,}\\
      (1 - \epsilon) \hat{A}_t, & \text{if $\hat{A}_t < 0$.}
    \end{cases}    
\label{eq:clip-function}
\end{equation}
The use of the clip function is to prevent the newly updated policy, i.e., $\Omega_{\boldsymbol{\theta}_{k}}$, being attracted to go far away from the old policy, i.e., $\Omega_{\boldsymbol{\theta}_{k-1}}$. In particular, the training of PPO is more stable because the gradient step update is clipped in the interval $[1 - \epsilon, 1 + \epsilon]$.
When the training objective function $L(\boldsymbol{\theta})$ is obtained, the parameter vectors of the actor and critic networks can be calculated as in lines 8 and 9 in Algorithm 1, respectively. The parameter vectors are iteratively updated until the cumulative reward converges to a stationary value~\cite{silver2014}.

\section{Performance Evaluation}
\label{sec:performance-evaluation}
\subsection{Simulation Settings}
\label{subsec:simulation-settings}

The details of our simulation settings are set as follows. For the channel gains between the users and the BS, we adopt the channel model in~\cite{mao2018}. In particular, the channel information at the BS contains the errors caused by quantized feedback from the users, i.e., imperfect channel state information. The number of antennas is set at $M = K = 6$ and the number of groups is set at $G = 3$. 
The constant channel gain is realized as $\mathbf{\bar{h}}_k = g_k \times [1, e^{j\phi_k}, e^{j2\phi_k}, e^{j3\phi_k}, e^{j4\phi_k}, e^{j5\phi_k}]$, where $g_k \in \mathbb{R}$ and $\phi_k \in [0, 2\pi]$ are control variables~\cite{mao2018}.
The feedback error of user $k$ at the BS follows a Complex Gaussian distribution with zero-mean and variance $\sigma_k^2 = g_k P_0^{-\beta_k}$ where $\beta_k$ is the degree of freedom (DoF) variable.

We utilize a dataset that consists of 88 360-degree videos with different categorized users' behaviours~\cite{dharmasiri2021}. The users' behaviours include the following features: (i) position of the HMDs, (ii) speed of the HMDs, (iii) maximum angle, and (iv) the area explored, i.e., the percentage coverage on the sphere by the $100^{\circ} \times 100^{\circ}$ FoV. The dataset usage is twofold, i.e., rendering FoVs for users and dividing users into groups based on their interests. 
We select one video for rendering FoVs with Vue-VR open-source software~\cite{ibrahim2022}. Fig.~\ref{fig:fov-render} illustrates a frame of the selected video before rendering and after rendering, respectively.

To evaluate the performance of the proposed learning approach, we use different baselines in this section. In particular, a multiple access baseline framework and two baseline algorithms are used as follows.

\begin{itemize}
\item SDMA is taken into consideration as a multiple access baseline scheme. In SDMA, the messages are separately encoded into different streams to the users. Each receiver decodes its message by treating interference from other streams as noise.
\item A Myopic Policy (MP) algorithm is adopted as a baseline algorithm.
In particular, the MP algorithm selects an action $a_t$ that maximizes the immediate reward signal $r_t$ at time step $t$ without concerning the long-term cumulative reward. To balance between exploration and exploitation, the action $a_t$ is randomly selected from the action space $\mathcal{A}$ if the previous action, i.e., $a_{t-1}$, results in a lower reward value than the highest historic reward value stored in the memory. Otherwise, the BS keeps selecting the recent action to exploit the benefit of getting high rewards.
\item A Uniform Power Allocation (UPA) is adopted as a baseline algorithm. UPA is widely used in RSMA optimization problems due to its low complexity~\cite{dai2016}. UPA allocates a fraction of the power to the common stream $z_c$ and equally divides the rest of the power to the other streams. Similarly, the rates of the streams can be allocated in the same way. As a result, the action space of the BS can be significantly reduced, e.g., from $40$ dimensions to $3$ dimensions in the case of $K=6$ users and $G=3$ groups. 
\end{itemize} 

In the training process, we train all the algorithms with $10^4$ samples from the environment. In the testing process, all the algorithms are run in $10^3$ samples that are independent and identically distributed (i.i.d) from the environment. In our setting, we consider the learning problem  episodic. Each episode is equivalent to a successful transmission of the 360-degree video to all users. We use the 360-degree video from~\cite{dharmasiri2021} which is publicly available. The video length is 3.26 minutes with different available qualities, i.e, from $360$ pixels to $4$K. The length of each time step is $2$ seconds and the episode terminates at time step $103^{th}$ (i.e., after 3.26 minutes). In the following, we evaluate the training and testing processes of the proposed DRL algorithm, compared to the baseline solutions. In particular, the results of the training process are illustrated in Fig.~\ref{fig:convergence-curves} and the results of the testing process are illustrated in Figs.~\ref{fig:power-varies},  \ref{fig:vid-varies}, \ref{fig:cpu-varies}, and~\ref{fig:cluster-varies}.
We select three main metrics that are important for both communication and computation functions of the BS: (i) average reward, (ii) system latency, and (iii) sum-rate. Note that we only consider the upper bound of system latency, i.e. $\max_{k \in \mathcal{K}}L_{s,k}$,  in the following evaluation. The reason is that we prioritize the fairness among all users in the system. 
Details of our simulation settings are summarized in Table~\ref{tab:simulation-settings}. In the following evaluation, we use the default settings as shown in Table~\ref{tab:simulation-settings}. Other parameters will be specified in particular simulations.

\begin{table}[t]
\begin{tabular}{c|l|l}
\hline \hline & \multicolumn{1}{c}{ \textbf{Communication Parameters} } & \textbf{Default Settings} \\
\hline$M$ & Number of antennas & $M=6$\\ 
\hline$K$ & Number of users & $K=6$\\
\hline$G$ & Number of groups & $G=3$~\cite{mao2018}\\
\hline$P_0$ & Power budget at the BS & $P_0 = 20$ dB\\
\hline$\mathbf{h}_k$ & Imperfect channel between  & $\mathbf{h}_k = \mathbf{\bar{h}}_k + \tilde{h}_k$\\
            & user $k$ and the BS        & \cite{mao2018, dai2016}\\
\hline$\mathbf{\bar{h}}_k$ & Constant channel gain & $\mathbf{\bar{h}}_k \in \mathbb{C}^{M \times 1}$ \\
\hline$\tilde{h}_k$ & Feedback error from users & $\tilde{h}_k \in \mathcal{CN}(0, \sigma_k^2)$ \\
\hline$\hat{h}_k$ & Root mean square of $\mathbf{h}_k$ & $\hat{h}_k \in \mathbb{R}$ \\
\hline$\sigma_k^2$ & Variance of feedback error & $\sigma_k^2 = g_k P_0^{-\beta_k}$ \\
\hline$\beta_k$ & Degree of freedom variable & $\beta_k = 0.6$ \\
\hline$N_x$ & Length of transmit signal $\mathbf{x}$ & $N_x = 0.1$ MB \\
\hline$B$ & Bandwidth & $B = 100$ MHz~\cite{mangiante2017}\\
\hline \hline \multicolumn{2}{c}{ \textbf{Computation Parameters} } & \textbf{Default Settings} \\
\hline$\eta$ & Computation capacity of the & $\eta = 2.3$ GHz\\
            &  BS        & \\
\hline & Video quality for the FoV  & $1280 \times 702$ \\
            & pre-rendering process & pixels\\
\hline & Dataset for users' behaviors & \cite{dharmasiri2021} \\
\hline
\end{tabular}
\caption{Parameter settings.}
\label{tab:simulation-settings}
\end{table}

\subsection{Simulation Results}
\subsubsection{Training and Testing}

\begin{figure}[h]
\centering
\includegraphics[width=0.9\linewidth]{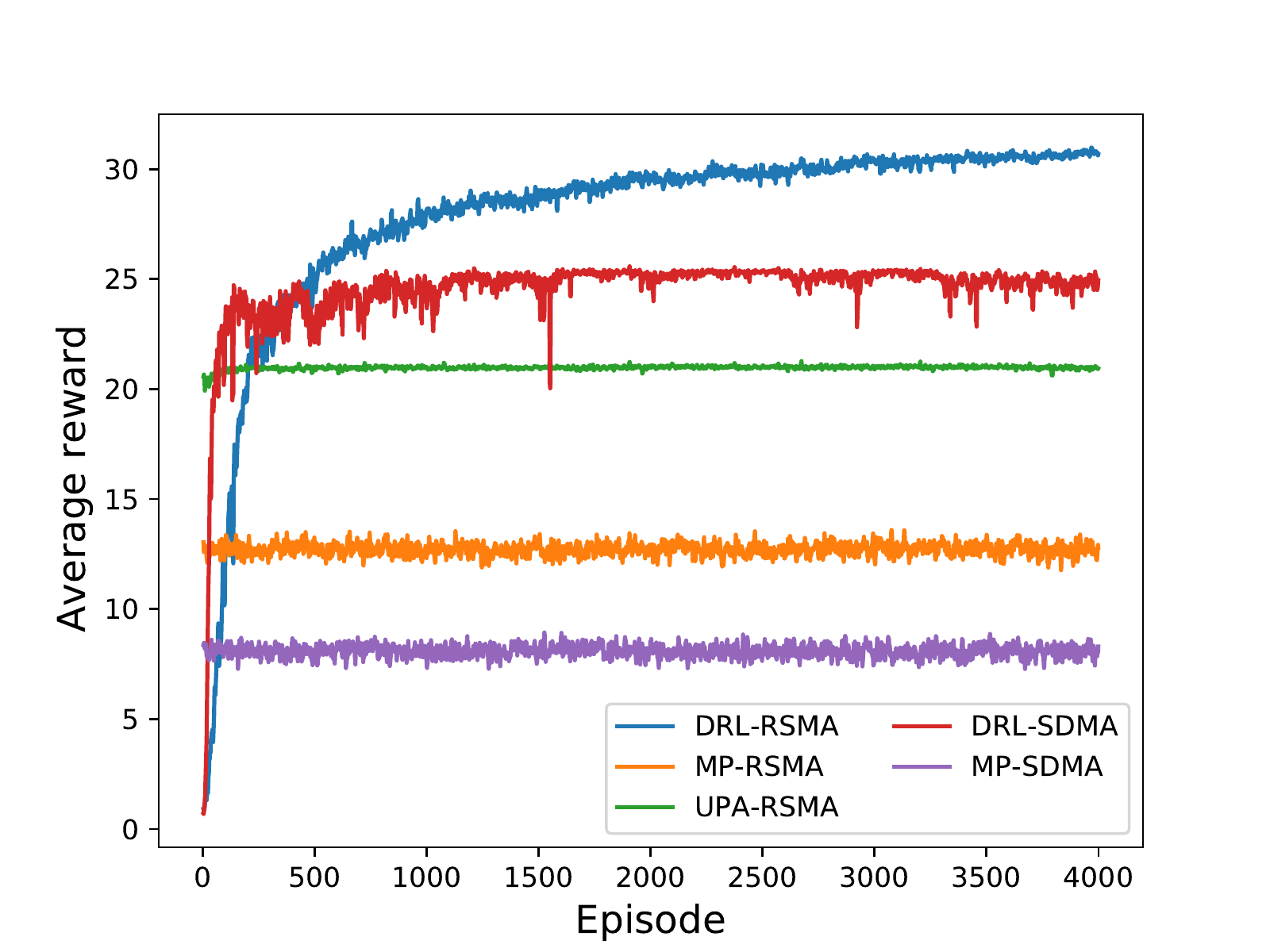}
\caption{Average rewards of our proposed approach, i.e., DRL-RSMA, compared to other approaches, in the training process.}
\label{fig:convergence-curves}
\end{figure}

Fig.~\ref{fig:convergence-curves} illustrates the performance of our proposed approach, i.e., DRL-RSMA, compared to other approaches, in the training process. 
Under the RSMA scheme, it can be observed that our proposed algorithm, denoted by DRL-RSMA, outperforms other baseline algorithms, i.e., MP-RSMA and UPA-RSMA, in term of average reward values. In particular, the DRL algorithm can achieve the performance that is equivalent to $146\%$ and $236\%$ of the performance of the UPA and MP algorithms, respectively.
Under the SDMA scheme, the DRL algorithm clearly shows better performance than that of the MP algorithm. The obtained rewards of both DRL and MP under SDMA scheme are lower than those of their RSMA counterparts, i.e., by $20\%$ and $35\%$, respectively.
The results confirm that the RSMA scheme is more effective, thanks to the rate splitting technique. In addition, the stable convergence curve in Fig.~\ref{fig:convergence-curves} shows that our proposed DRL-RSMA scheme can not only deal with the large continuous action space, i.e., 40 dimensions in the default setting, but also the multimodal data from difference sources, i.e., imperfect channel state information, CPU usage of the BS, and users' behaviours.

\begin{figure*}[h]
	\centering
	\begin{subfigure}[b]{0.33\linewidth}
		\centering
		\includegraphics[width=1.1\linewidth]{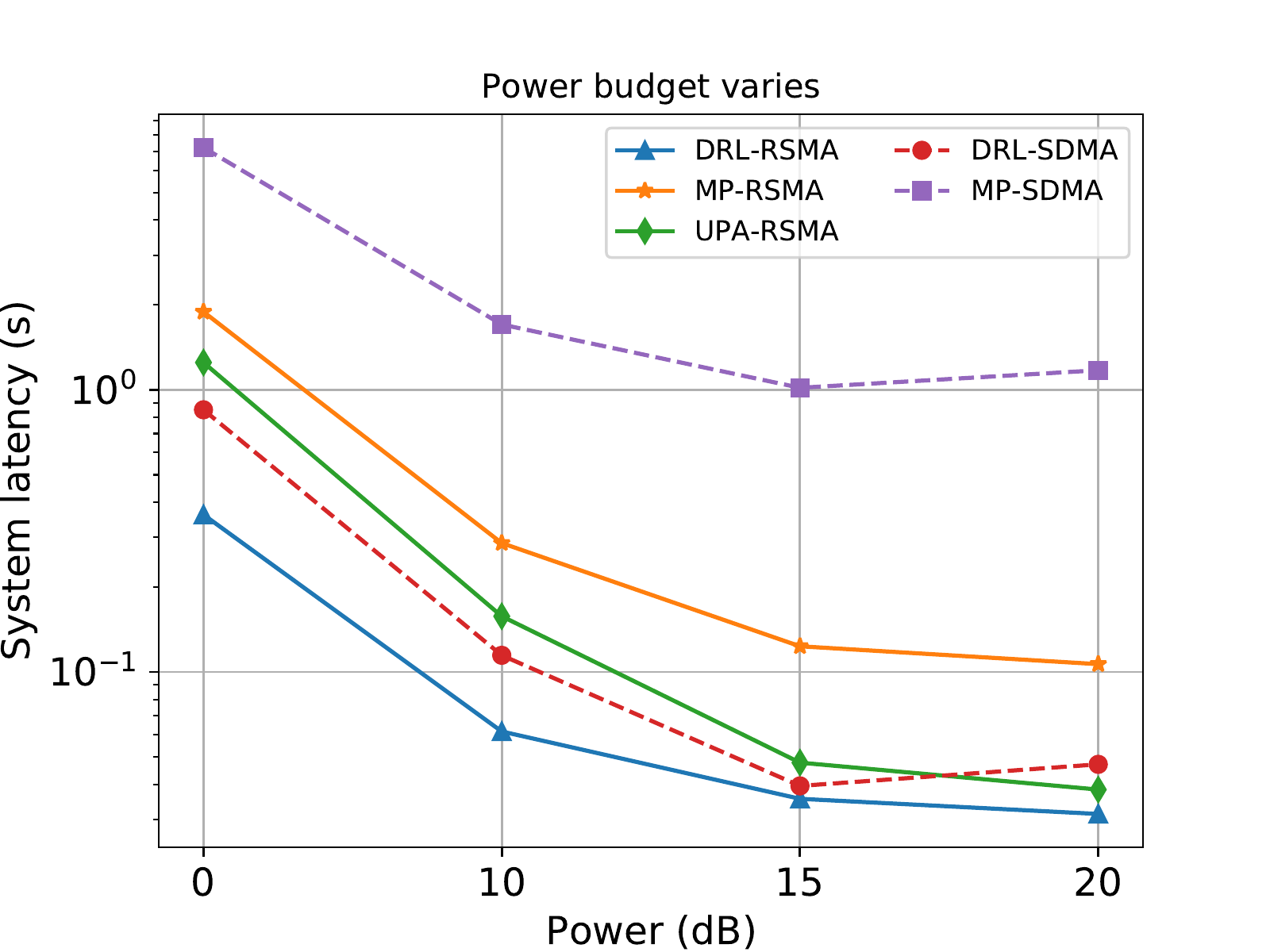}
		\caption{}
	\end{subfigure}%
	~ 
	\begin{subfigure}[b]{0.33\linewidth}
		\centering
		\includegraphics[width=1.1\linewidth]{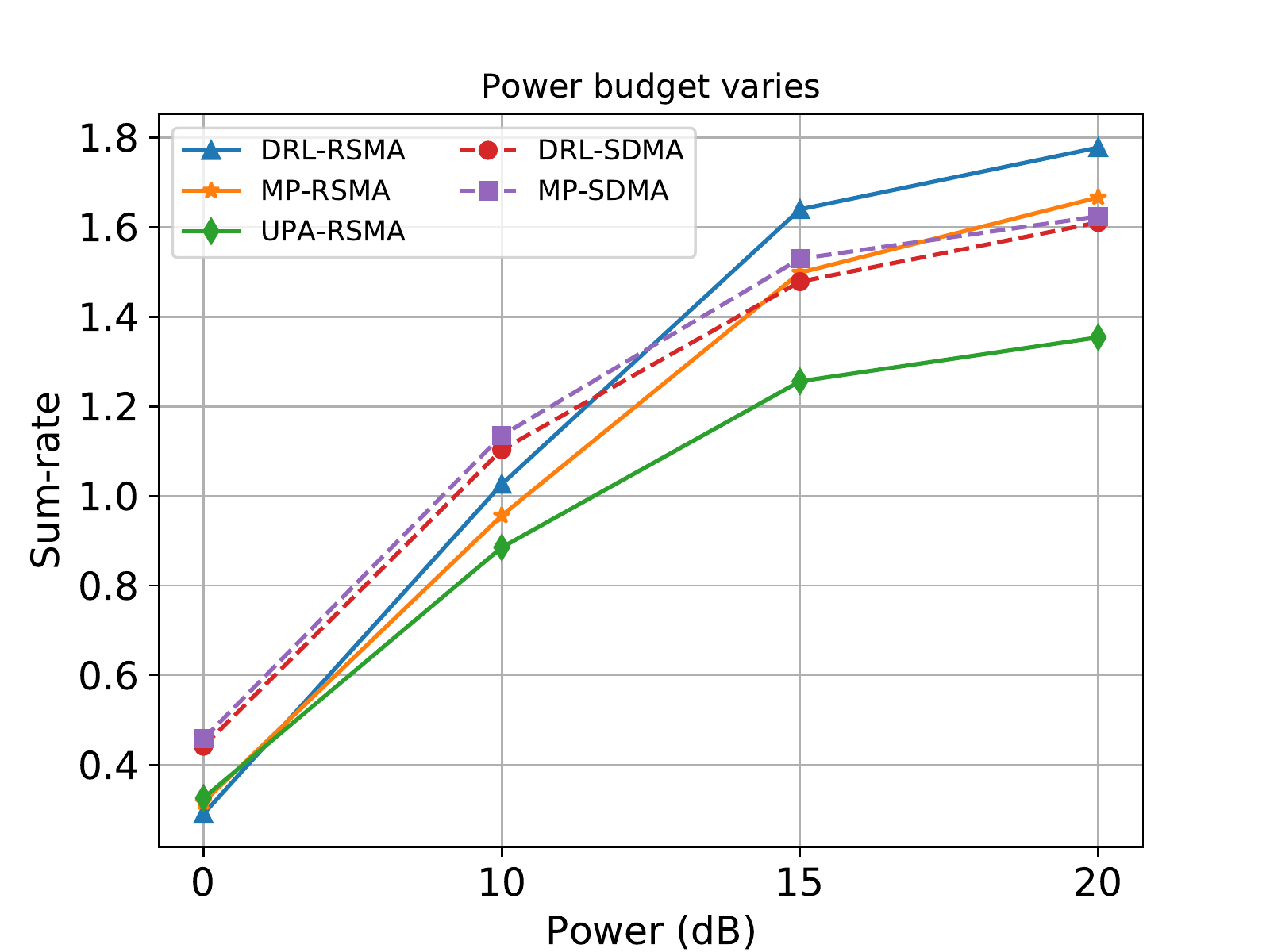}
		\caption{}
	\end{subfigure}%
	~
	\begin{subfigure}[b]{0.33\linewidth}
		\centering
		\includegraphics[width=1.1\linewidth]{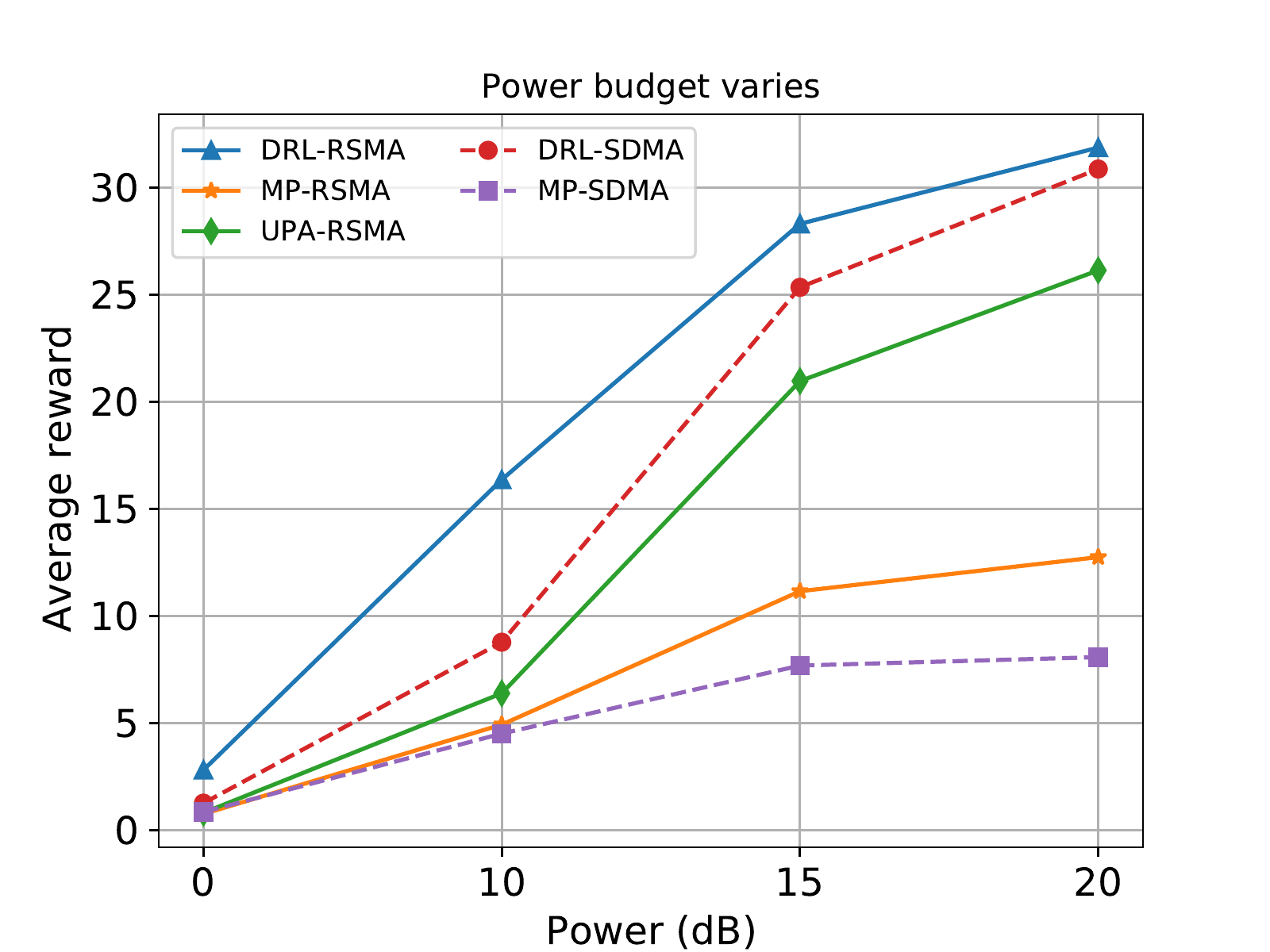}
		\caption{}
	\end{subfigure}%
	\caption{(a) System latency, (b) sum-rate, and (c) average reward values when the power budget at the BS varies.} 
	\label{fig:power-varies}
\end{figure*}
\begin{figure*}[h]
	\centering
	\begin{subfigure}[b]{0.33\linewidth}
		\centering
		\includegraphics[width=1.1\linewidth]{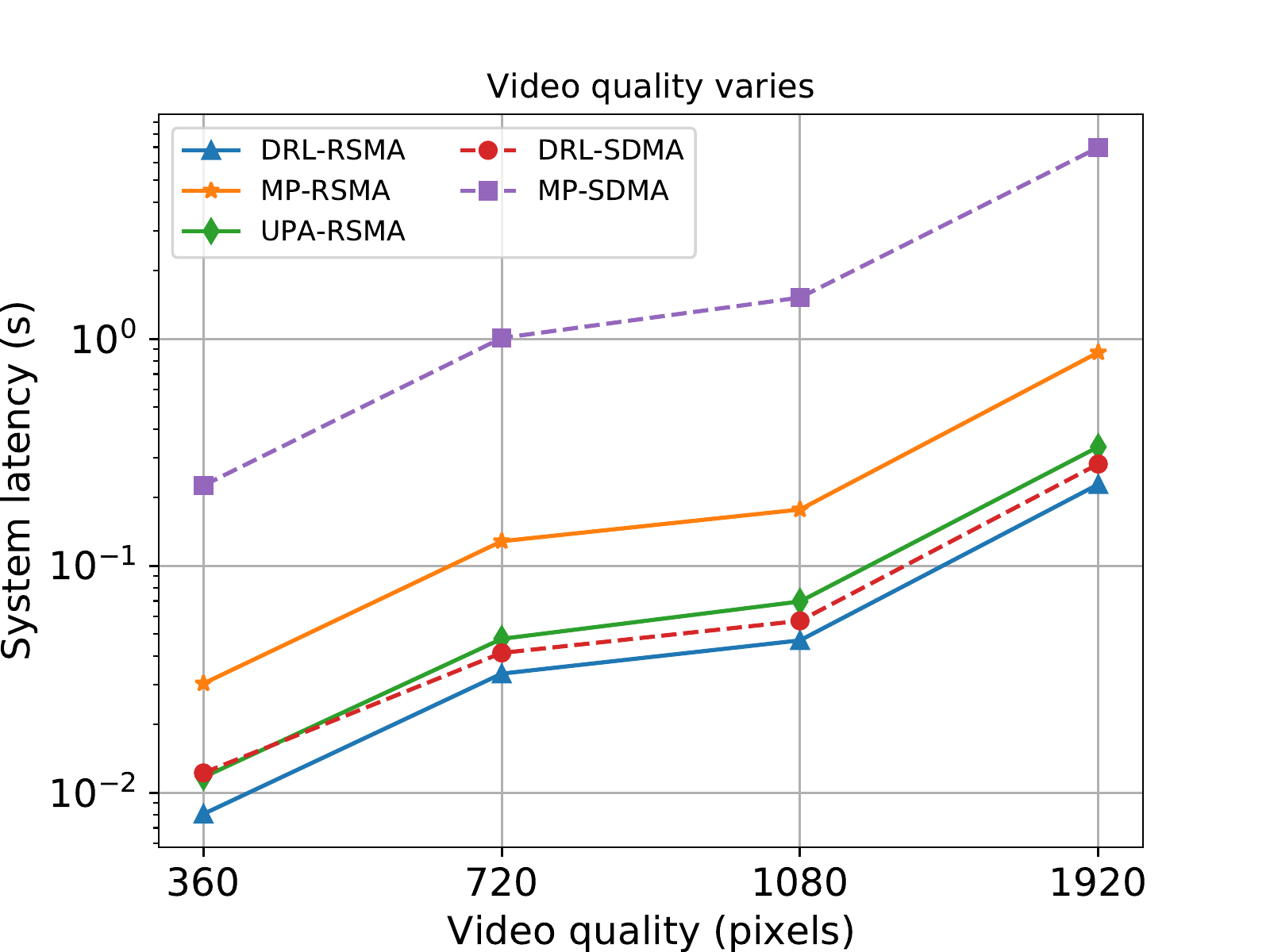}
		\caption{}
	\end{subfigure}%
	~ 
	\begin{subfigure}[b]{0.33\linewidth}
		\centering
		\includegraphics[width=1.1\linewidth]{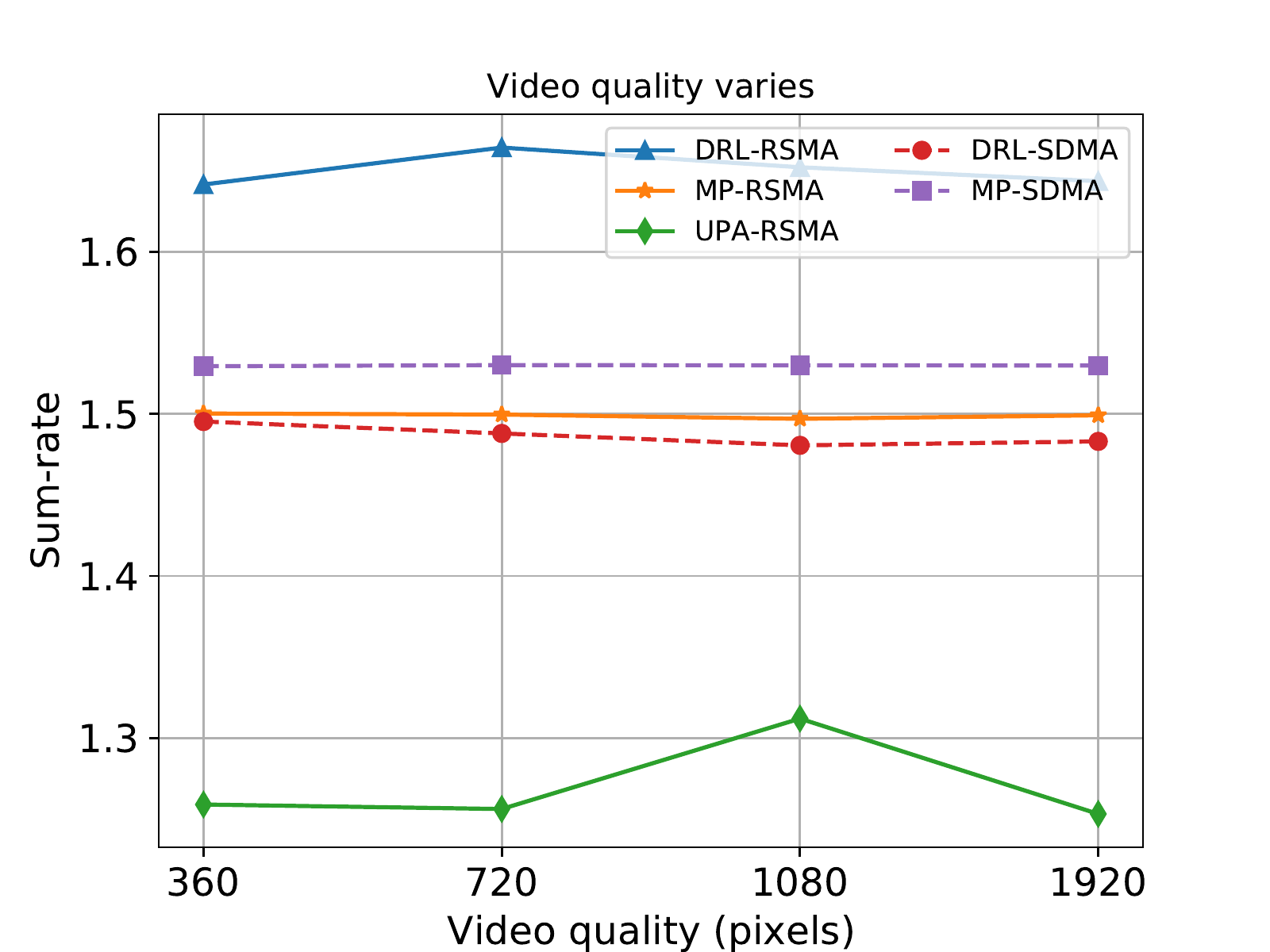}
		\caption{}
	\end{subfigure}%
	~
	\begin{subfigure}[b]{0.33\linewidth}
		\centering
		\includegraphics[width=1.1\linewidth]{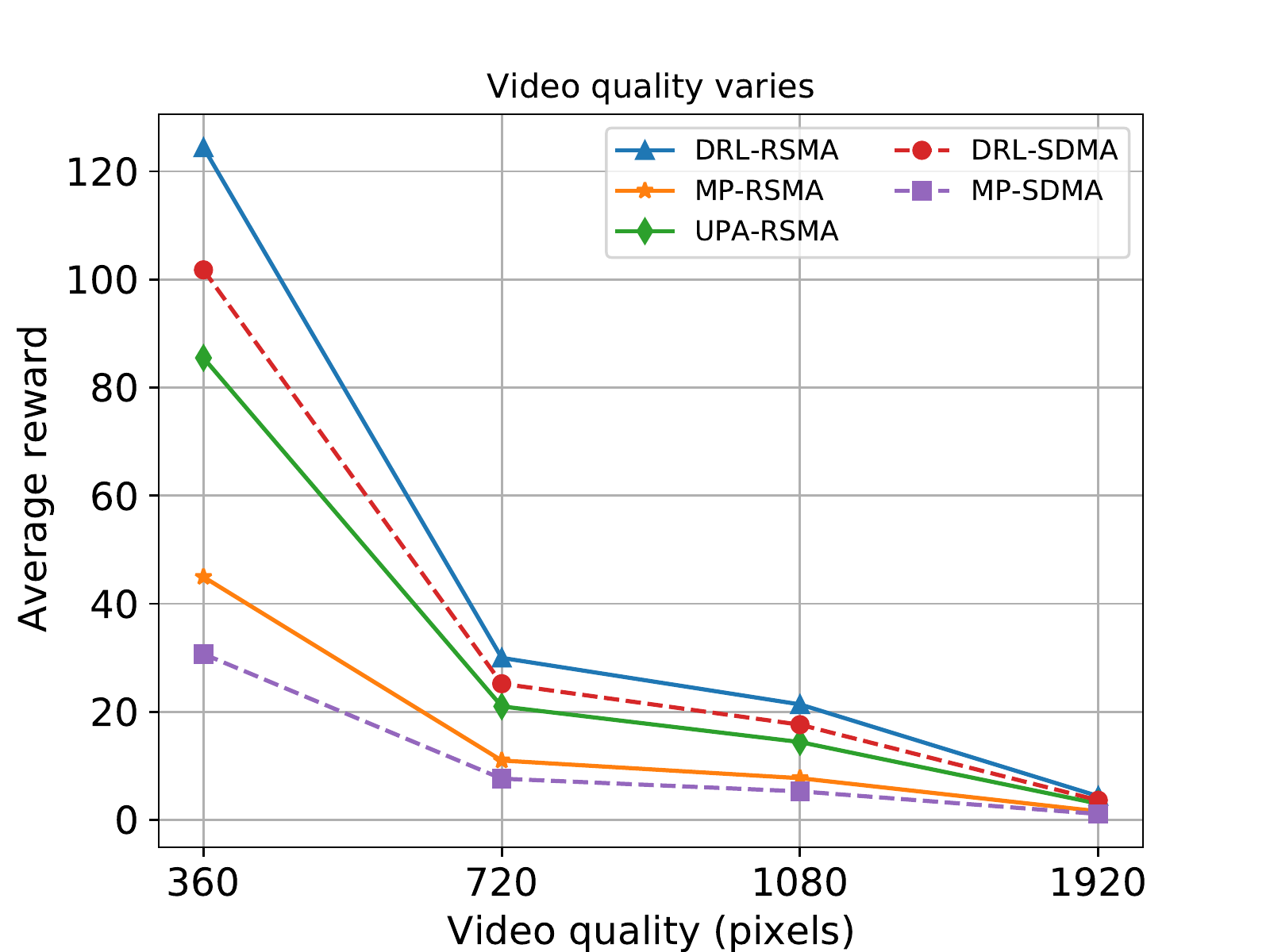}
		\caption{}
	\end{subfigure}
	\caption{(a) System latency, (b) sum-rate, and (c) average reward values when the video quality for FoV pre-rendering at the BS varies.} 
	\label{fig:vid-varies}
\end{figure*}

In Fig.~\ref{fig:power-varies}, we vary the power budget, i.e. $P_0$, at the BS from $0$ dB to $20$ dB to evaluate the performance of the algorithms.
In Fig.~\ref{fig:power-varies}(a), we can observe that the proposed DRL-RSMA scheme can achieve the lowest latency for the system.
Specifically, when the BS's power is at $10$, $15$, and $20$ dB, the system latency can achieve order of magnitude $10^{-3}$ second, i.e.,  $61$, $35$, and $31$ milliseconds, respectively. These values of latency can guarantee smooth experiences for the users~\cite{mangiante2017}. With MP-SDMA and MP-RSMA approaches, the latency values are greater than $100$ milliseconds, which may cause motion sickness for users~\cite{mangiante2017}.
Furthermore, in Fig.~\ref{fig:power-varies}(b), it can be observed that all the sum-rate values increase as the power budget at the BS increases and the sum-rate values of the proposed DRL-RSMA approach are higher than those of other approaches with power values are $15$ and $20$ dB. The reason is that the RSMA framework is more effective in high-power values~\cite{mao2018}. It is also noted that the sum-rate values obtained in our simulations are lower than those in~\cite{mao2018, hieu2021} and~\cite{hieu2022} because our main goal is to minimize the upper bound of system latency to guarantee the user fairness, rather than maximizing the sum-rate of the system.
In Fig.~\ref{fig:power-varies}(c), we can observe that all the reward values increase as the power budget at the BS increases, and the proposed DRL-RSMA scheme achieves the highest reward values, followed by DRL-SDMA, UPA-RSMA, MP-RSMA, and MP-SDMA, respectively. 

\begin{figure*}[h]
	\centering
	\begin{subfigure}[b]{0.33\linewidth}
		\centering
		\includegraphics[width=1.1\linewidth]{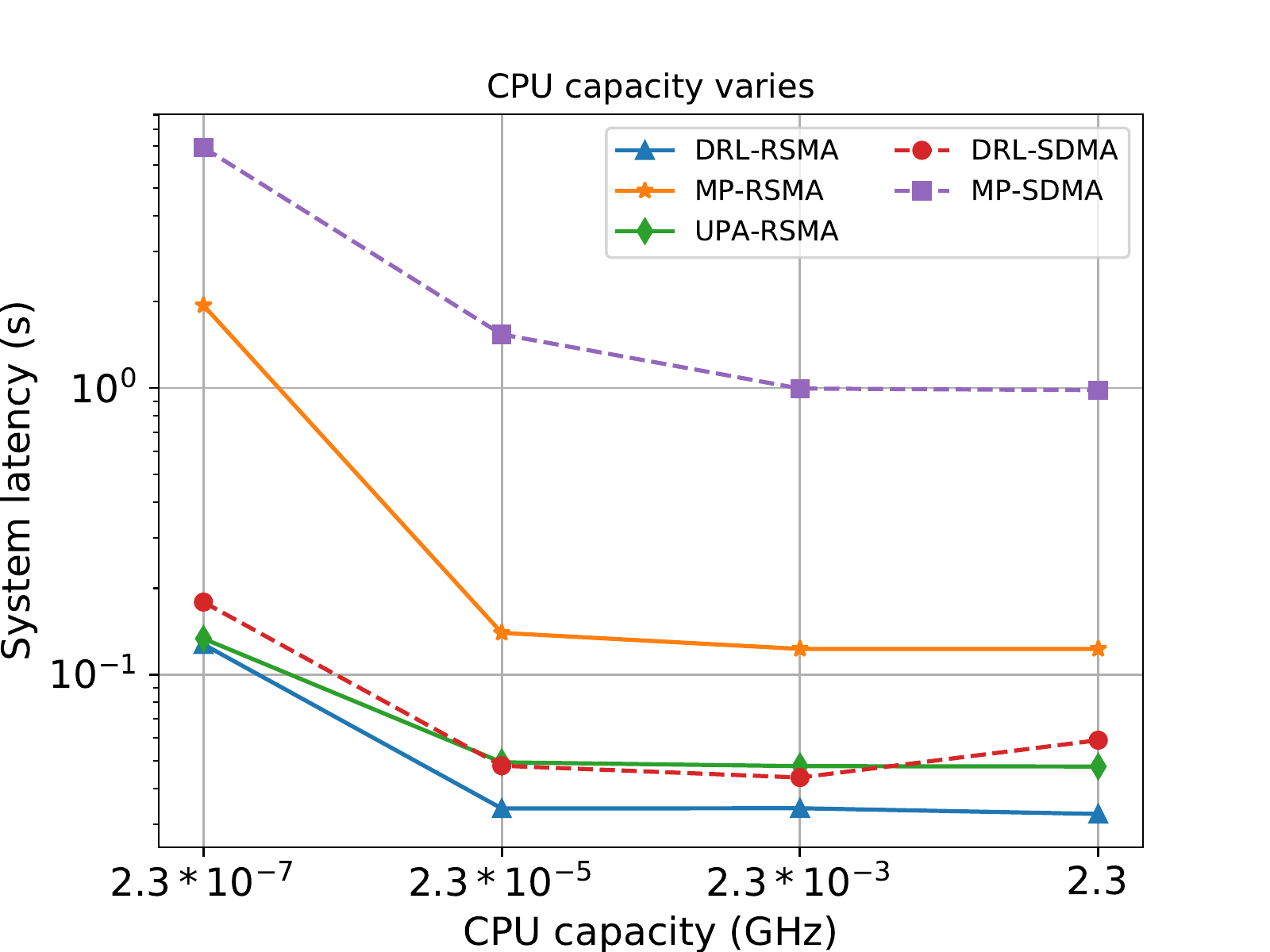}
		\caption{}
	\end{subfigure}%
	~ 
	\begin{subfigure}[b]{0.33\linewidth}
		\centering
		\includegraphics[width=1.1\linewidth]{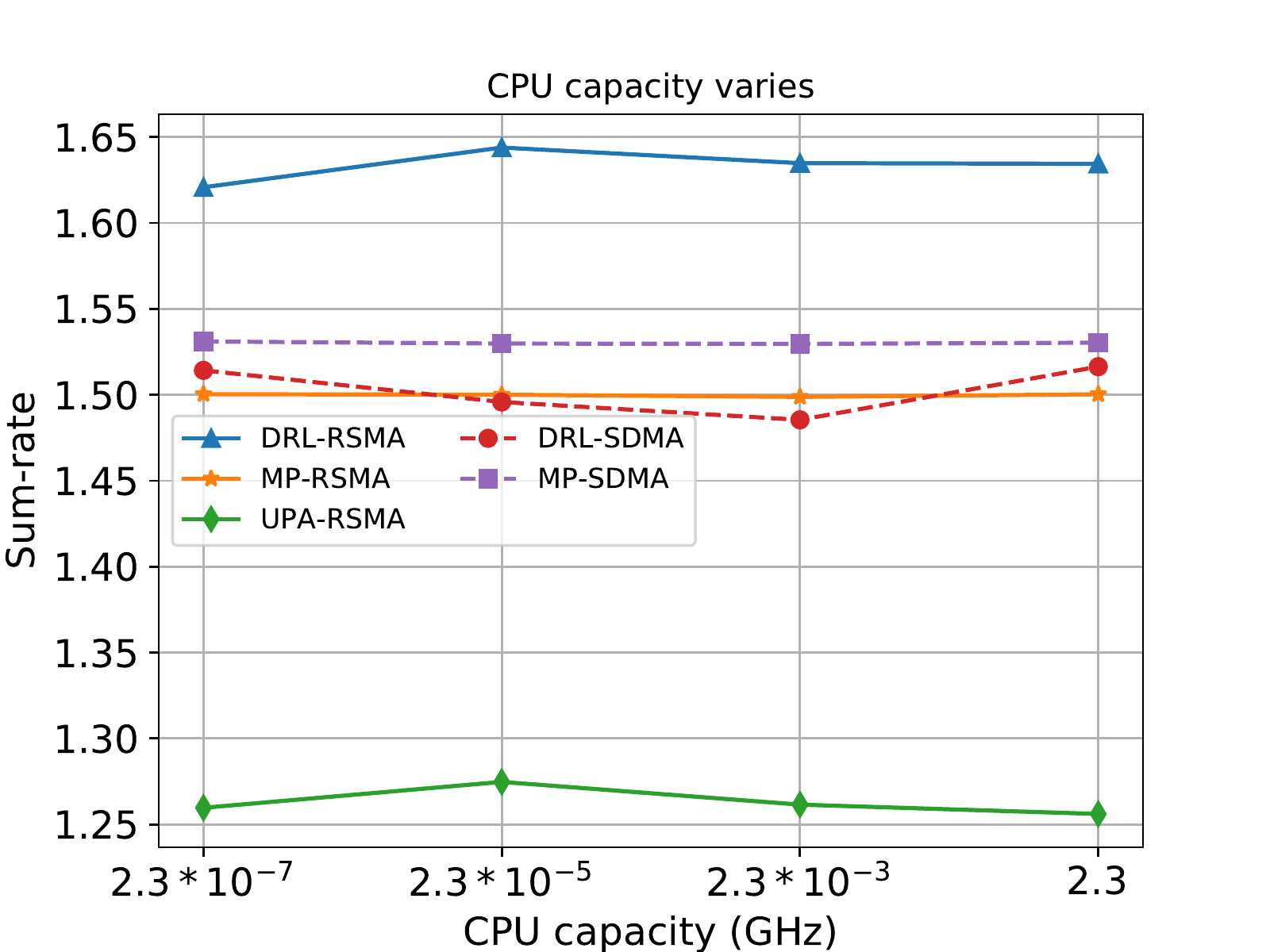}
		\caption{}
	\end{subfigure}%
	~
	\begin{subfigure}[b]{0.33\linewidth}
		\centering
		\includegraphics[width=1.1\linewidth]{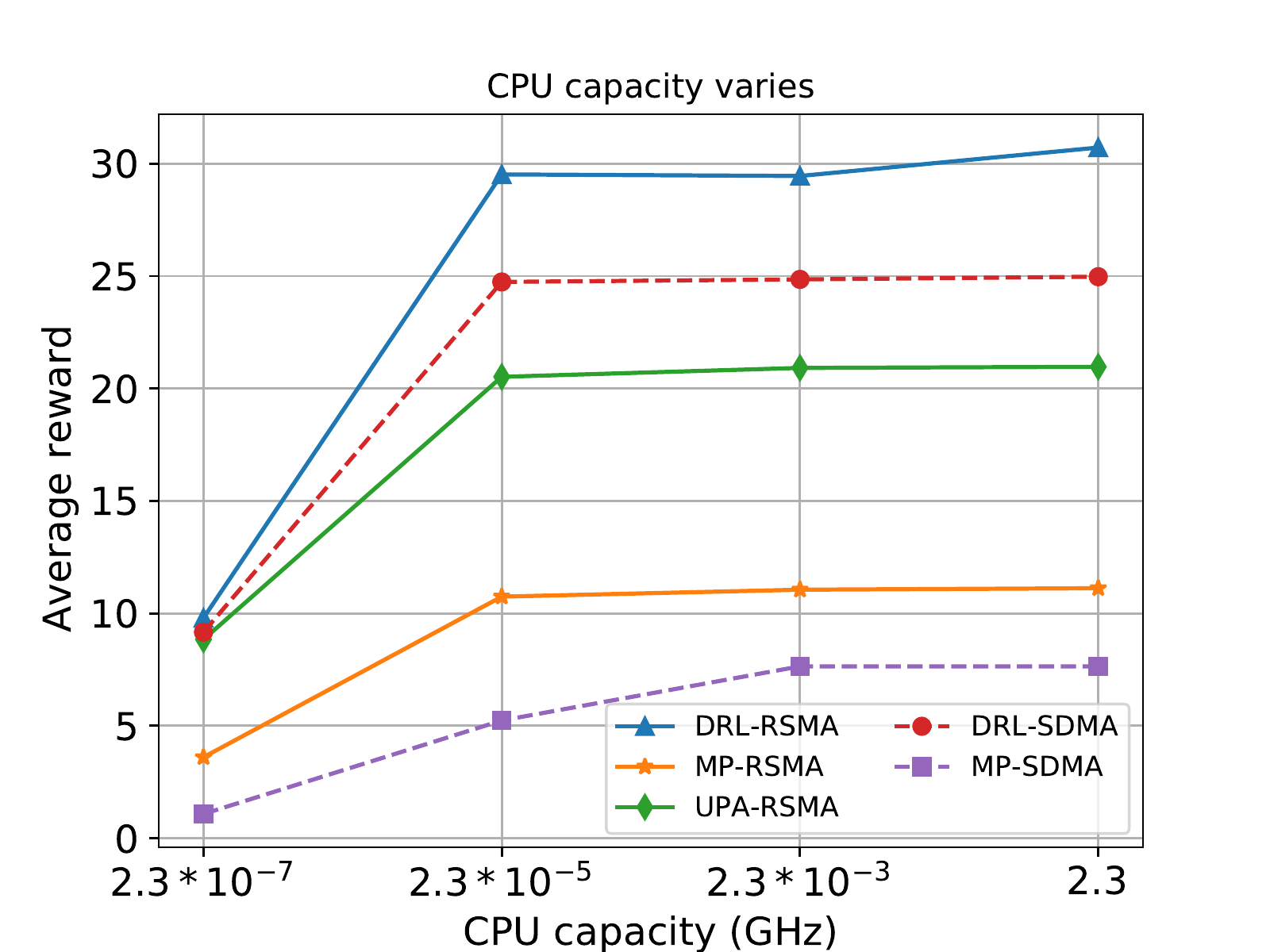}
		\caption{}
	\end{subfigure}%
	\caption{(a) System latency, (b) sum-rate, and (c) average reward values when the CPU capacity at the BS varies.} 
	\label{fig:cpu-varies}
\end{figure*}

Next, we vary the quality of the 360-degree video that is used for the FoV pre-rendering process at the BS. In Fig.~\ref{fig:vid-varies}, we increase the video quality from $640 \times 360$ pixels, denoted as $360$ for short, to $3840 \times 1920$ pixels, denoted as $1920$ for short. 
Fig.~\ref{fig:vid-varies}(a) shows that the overall latency values obtained by all algorithms increase as the video quality increases. 
The reason is that the increase of video quality results in (i) increasing the FoV pre-rendering time at the BS and (ii) increasing the packet's length to be transmitted at each time step. 
It can be observed from Fig.~\ref{fig:vid-varies}(a) that our proposed DRL-RSMA approach can achieve order of magnitude  $10^{-3}$ for the system latency with video qualities $640 \times 360$ pixels, $1280 \times 720$ pixels, and $1920 \times 1080$ pixels, which will ensure smooth experiences for the users~\cite{mangiante2017}. 
For the $3840 \times 1920$ pixels video, the latency is still higher than $100$ milliseconds. Potential solutions to reduce the latency of using high quality video can be (i) increasing the power at the BS or (ii) decreasing the transmitted FoV frame sizes to the users. 
Increasing the power at the BS is a more straightforward solution that may require better power splitter circuit at the BS~\cite{krikidis2014}. Decreasing the transmitted FoV frame sizes may require clustering the users into groups that have smaller regions of interest. For example, the users can be divided into smaller groups that share the high overlapped FoVs. As such, the BS needs to transmit the FoVs having narrower frames, e.g., $80^{\circ} \times 80^{\circ}$ instead of $100^{\circ} \times 100^{\circ}$ in our setting. This approach does not require hardware upgrade or replacement at the BS but the clustering algorithms need to be highly accurate. We believe that this is also a potential direction which is worth investigated in future work. 
In Fig.~\ref{fig:vid-varies}(b), the sum-rate values remain the same under different values of video quality. 
The reason is that the sum-rate is only affected by the power budget at the BS. Fig.~\ref{fig:vid-varies}(c) shows that the average reward values of all algorithms decrease as the video quality increases, and our proposed DRL-RSMA approach obtains the highest reward values under all the considered scenarios. The average reward results show that our proposed algorithm guarantees the good convergence property as analysed in Fig.~\ref{fig:convergence-curves}.

To further evaluate the impacts of computational capacity of the BS to the system performance, we vary the CPU capacity of the BS as shown in Fig.~\ref{fig:cpu-varies}. 
Fig.~\ref{fig:cpu-varies}(a) shows that the latency values obtained by all algorithms decrease as the CPU capacity of the BS increases from $2.3 \times 10^{-7}$ GHz to $2.3$ GHz and our proposed DRL-RSMA appoach always achieves the best performance. The latency of our proposed DRL-RSMA approach meets the requirement of order of magnitude $10^{-3}$ second at most of the CPU capacity values, except $2.3 \times 10^{-7}$ GHz which is a relatively small computational unit. 
This implies that our proposed approach is highly efficient in terms of computational usage. 
Similar to above analysis, the sum-rate values in Fig.~\ref{fig:cpu-varies}(b) remain unchanged or slightly changed when the CPU capacity is varied.
In Fig.~\ref{fig:cpu-varies}(c), it can be observed that the average reward values increase as the CPU capacity of the BS increases, and our proposed DRL-RSMA approach always achieves the best performance under all scenarios.

\subsubsection{Benefits of Clustering Users based on FoVs}

\begin{figure*}[t]
	\centering
	\begin{subfigure}[b]{0.33\linewidth}
		\centering
		\includegraphics[width=1.0\linewidth]{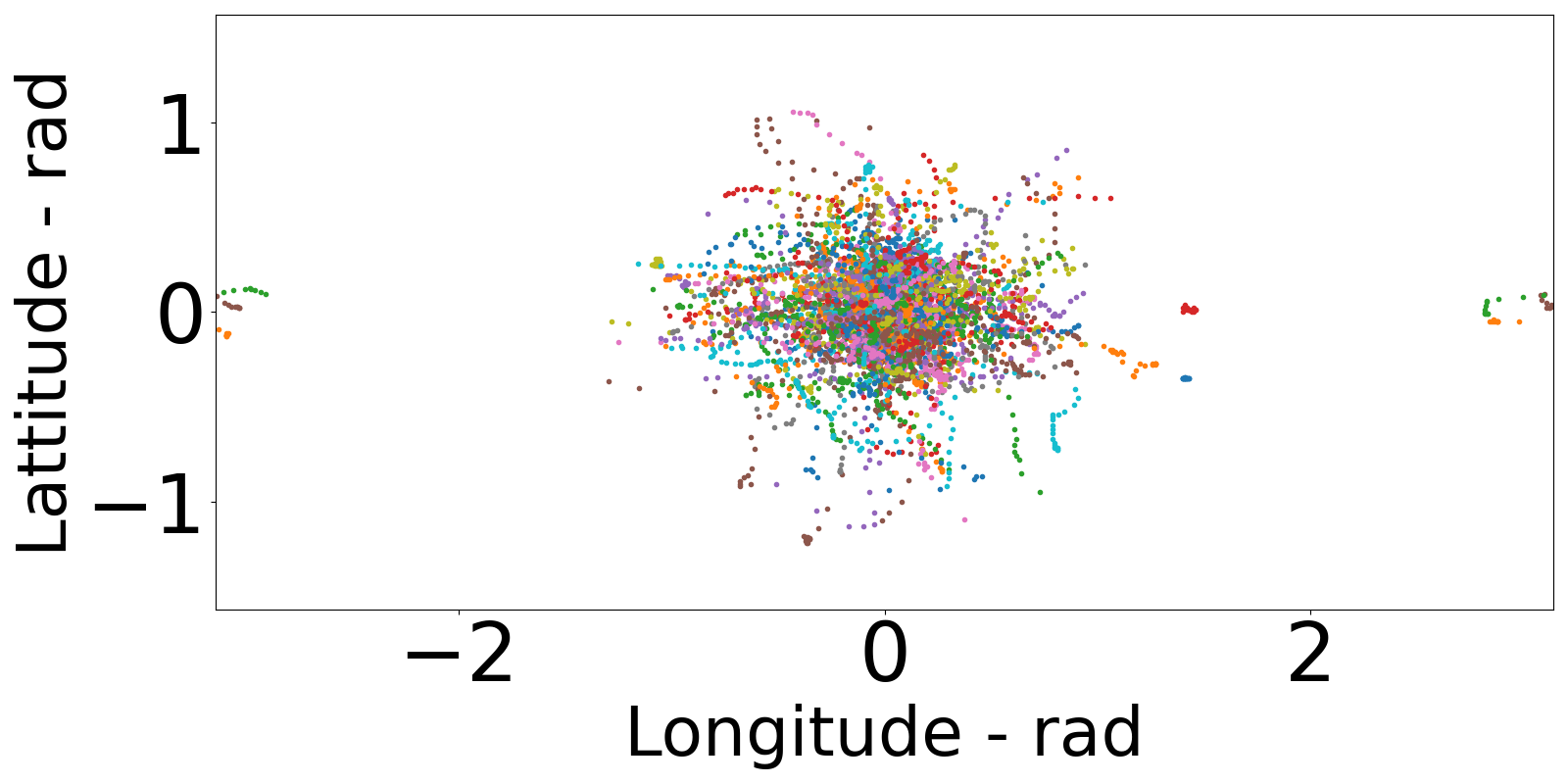}
		\caption{}
	\end{subfigure}%
	~ 
	\begin{subfigure}[b]{0.33\linewidth}
		\centering
		\includegraphics[width=1.0\linewidth]{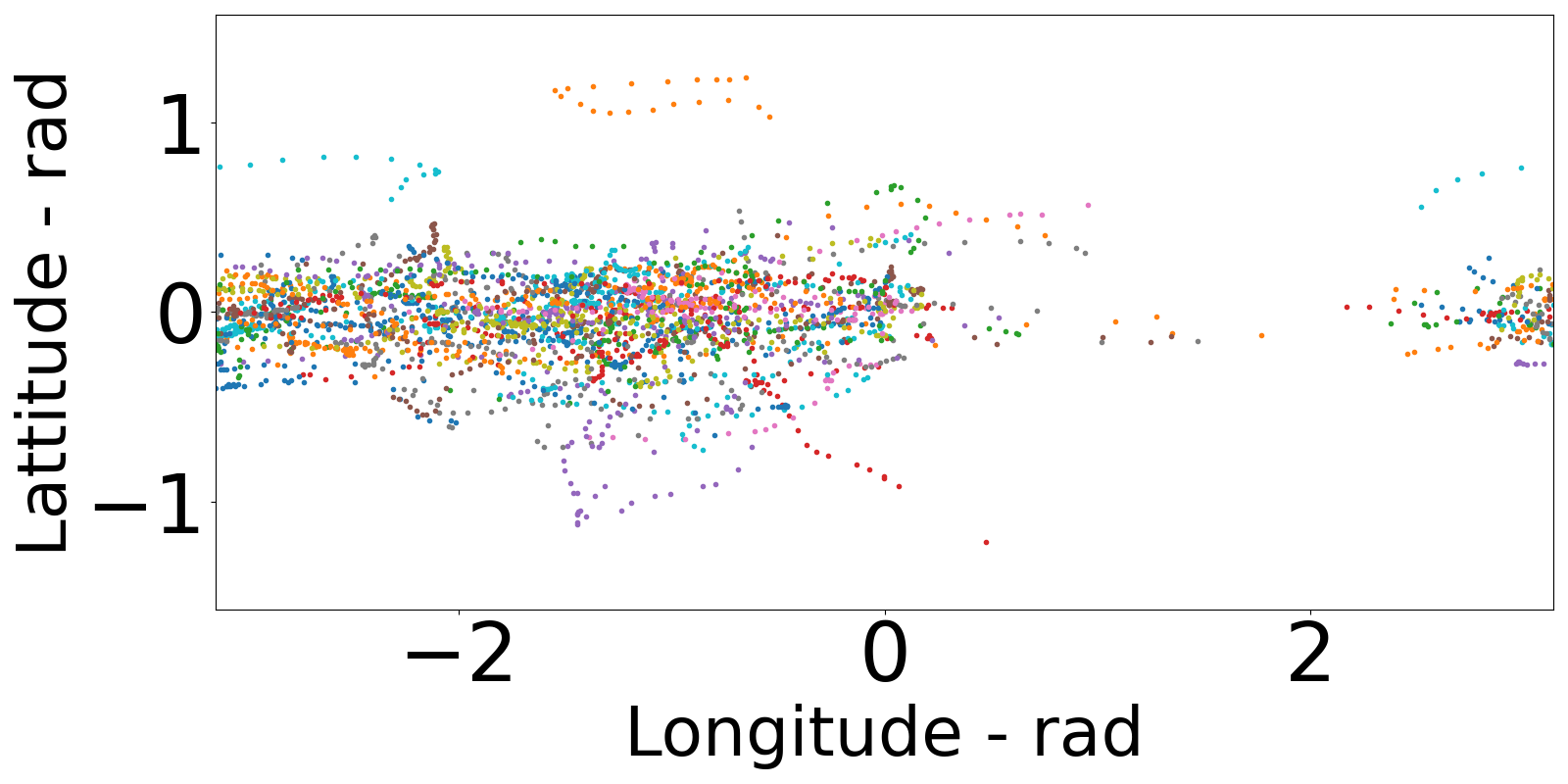}
		\caption{}
	\end{subfigure}%
	~ 
	\begin{subfigure}[b]{0.33\linewidth}
		\centering
		\includegraphics[width=1.0\linewidth]{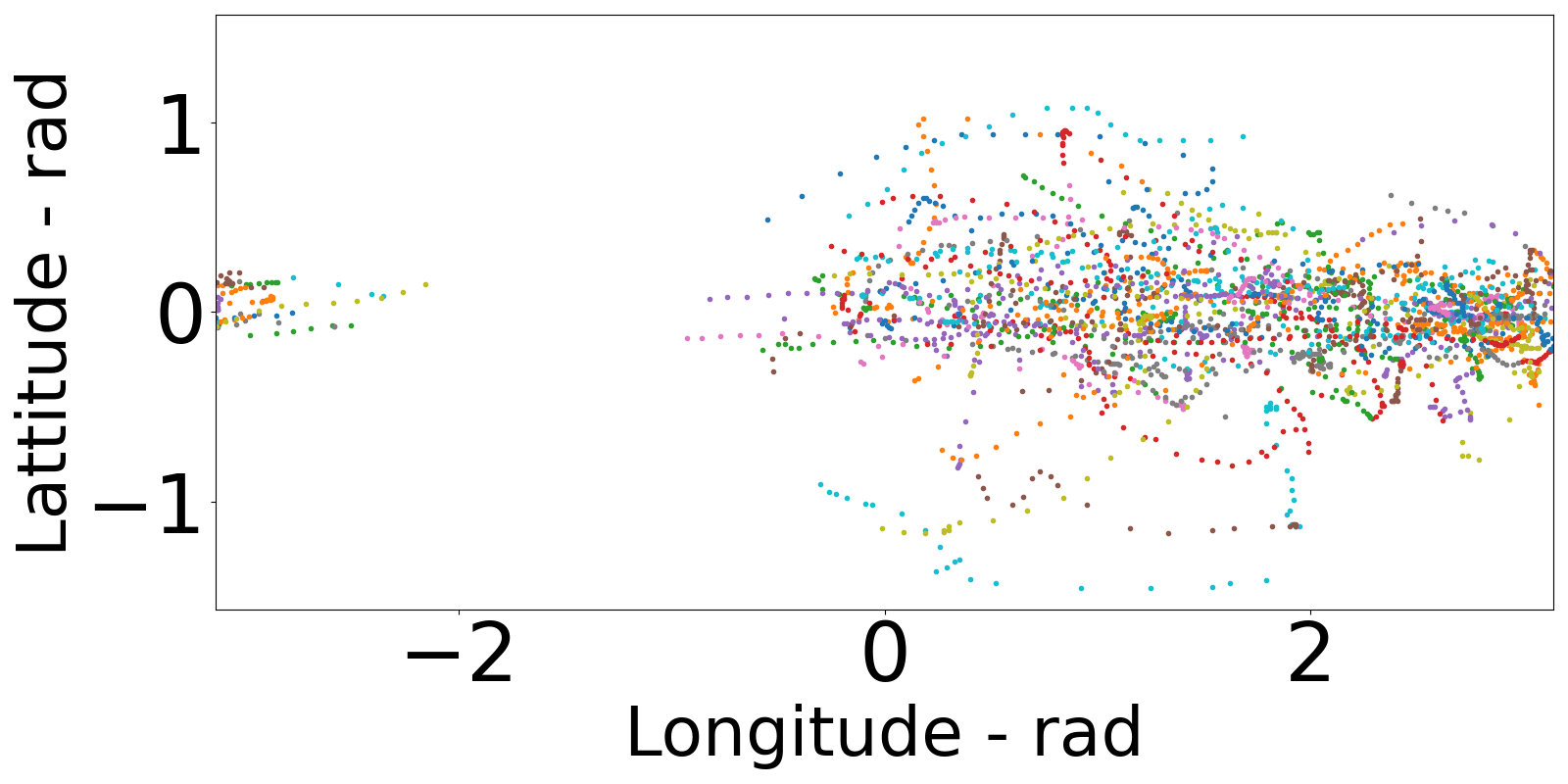}
		\caption{}
	\end{subfigure}%
	\caption{User clustering results for (a) the first group, (b) the second group, and (c) the third group. The 360-degree frames are first transformed to 2D frames. The scattered points are the central view points of the users's FoVs. Based on the categorized features in~\cite{dharmasiri2021}, three clusters are formed with the K-means cluster algorithm.
	} 
	\label{fig:vp-cluster}
\end{figure*}

\begin{figure*}[t]
	\centering
	\begin{subfigure}[b]{0.33\linewidth}
		\centering
		\includegraphics[width=1.0\linewidth]{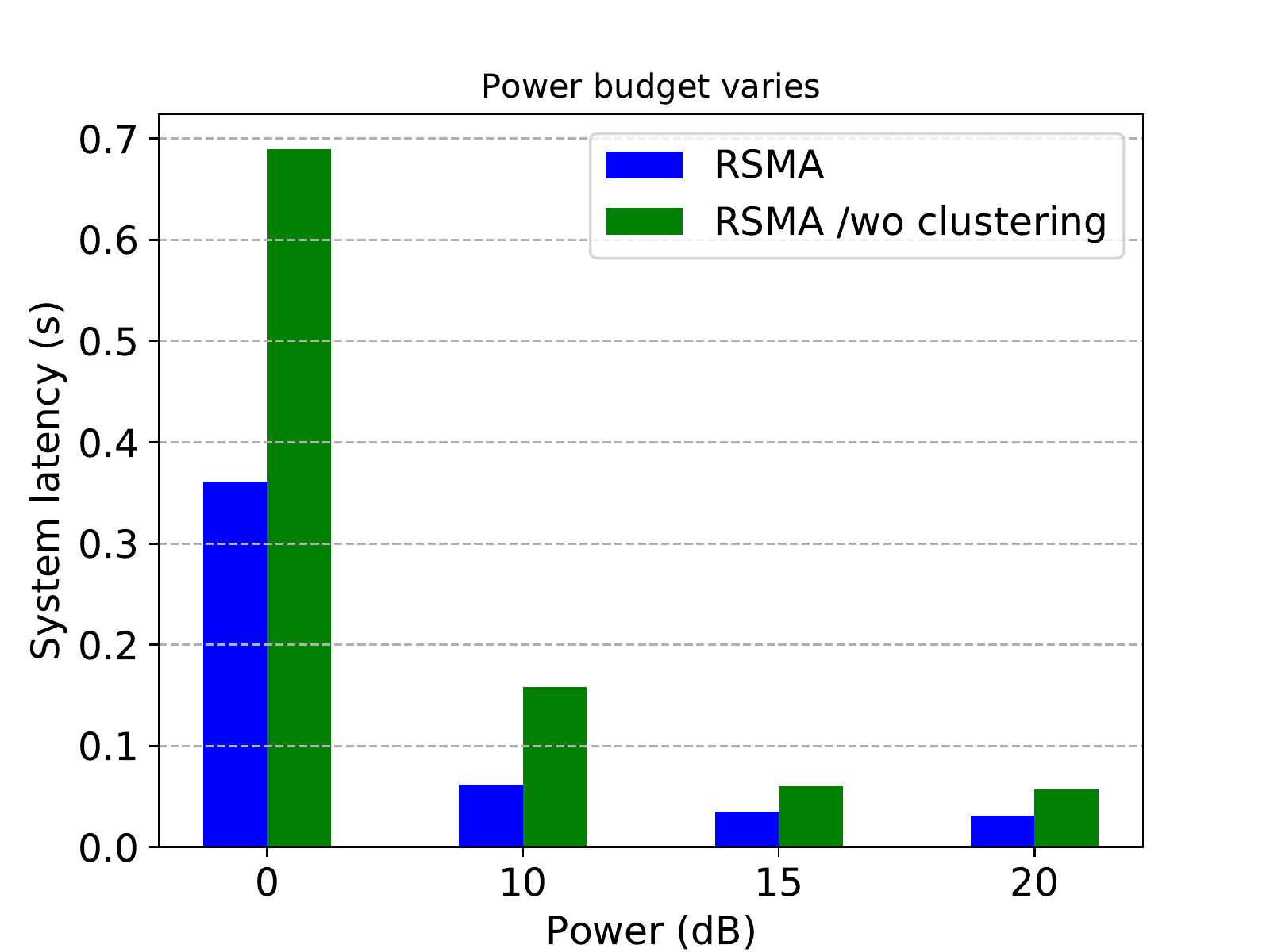}
		\caption{}
	\end{subfigure}%
	~ 
	\begin{subfigure}[b]{0.33\linewidth}
		\centering
		\includegraphics[width=1.0\linewidth]{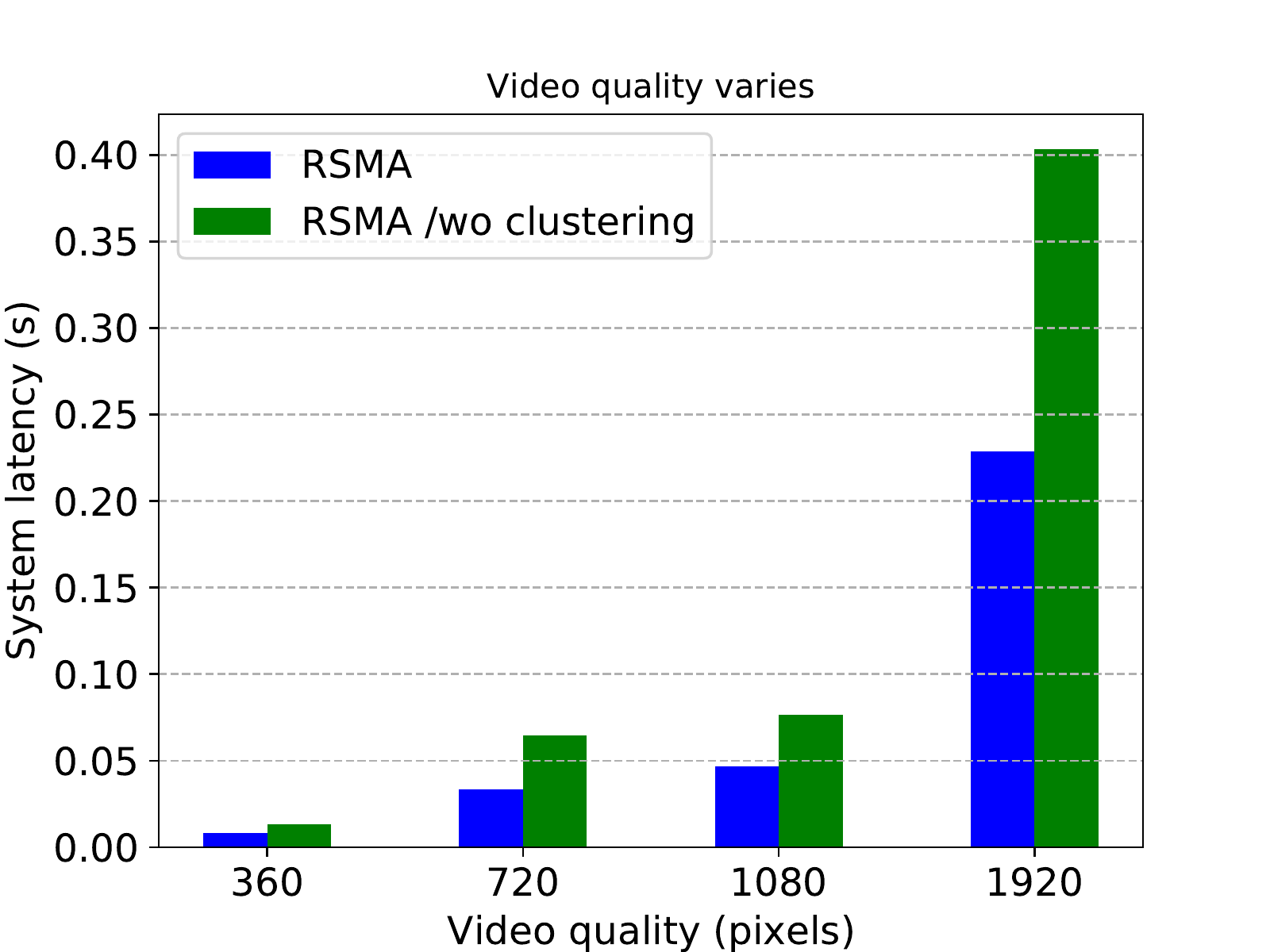}
		\caption{}
	\end{subfigure}%
	~ 
	\begin{subfigure}[b]{0.33\linewidth}
		\centering
		\includegraphics[width=1.0\linewidth]{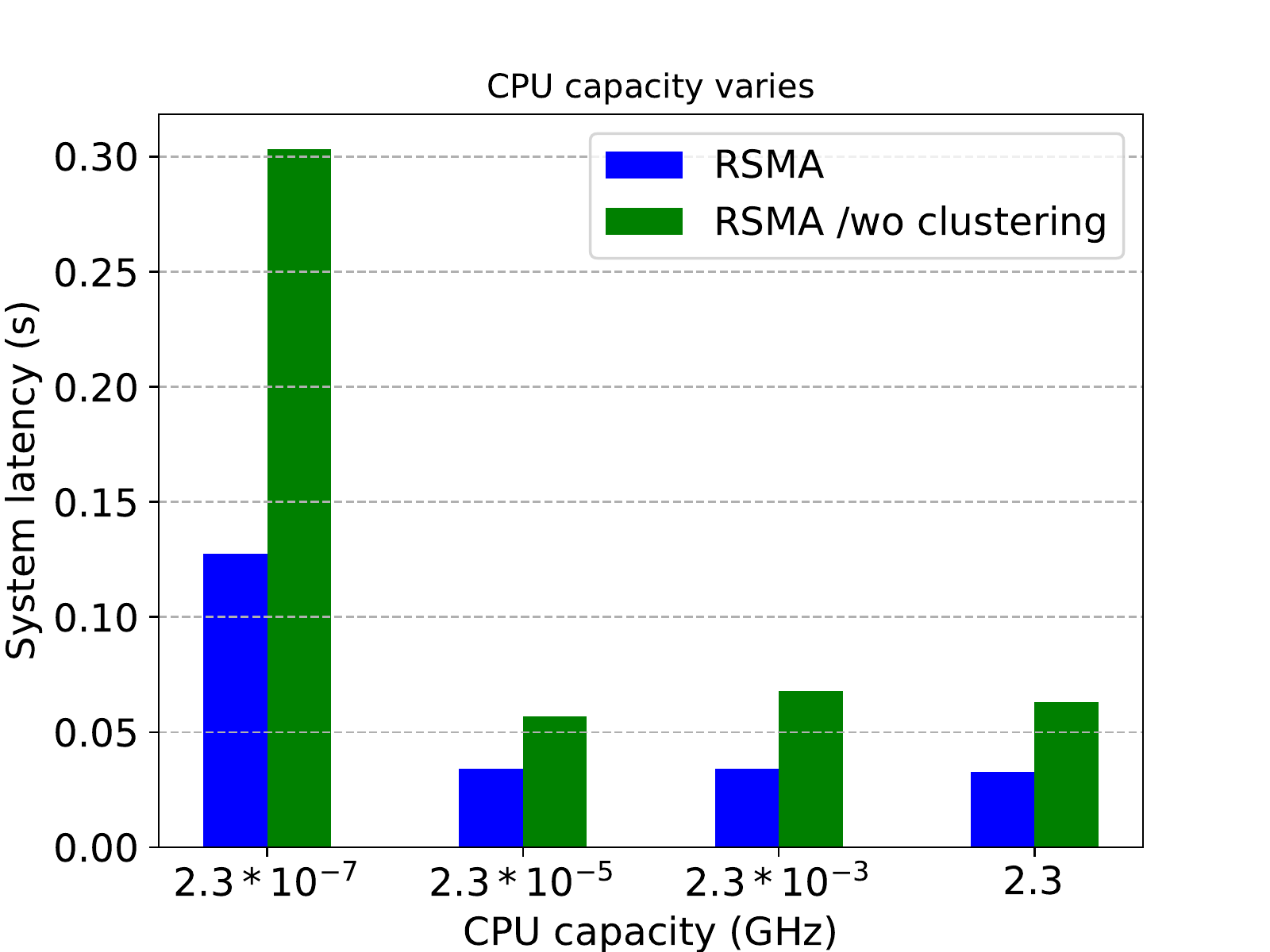}
		\caption{}
	\end{subfigure}%
	\caption{System latency of the DRL-RSMA approach with user clustering and DRL-RSMA approach without user clustering.} 
	\label{fig:cluster-varies}
\end{figure*}
As mentioned above, clustering users into different groups based on their FoVs can be beneficial by reducing the sizes of frames that need to be transmitted to the users. In the following, we evaluate the benefits of clustering users based on FoV and utilizing  HRSMA framework in Section~\ref{sec:system-model}.
In Fig.~\ref{fig:vp-cluster}, we show the results of the K-means cluster algorithm for $6$ users with the dataset from~\cite{dharmasiri2021}. Fig.~\ref{fig:vp-cluster} shows the scattered plots of central view points of users' FoVs.
It can be observed that the users can change their groups dynamically based on their interest. The users in the first group are more interested to watch the content at the center of the video, while the users in the second and third groups are interested to watch the contents at the left and right corners of the video, respectively. As a result, the BS only needs to transmit the FoVs that are around the center of the 360-degree video frame, i.e., $100^{\circ} \times 100^{\circ}$ in our setting, for the first group in Fig.~\ref{fig:vp-cluster}(a), instead of transmitting the whole $360^{\circ} \times 360^{\circ}$ frame.

Based on the above clustering results, the HRSMA scheme splits and transmits messages to groups of users in different power levels and rates. In Fig.~\ref{fig:cluster-varies}, the system latency values obtained by the proposed DRL-RSMA approach with clustering, denoted as ``RSMA", are lower than those of the DRL-RSMA approach without clustering, i.e., all the users are in one group, denoted as ``RSMA /wo clustering", in all scenarios. There are two main reasons for this results. First, dividing users into groups can achieve better rate performance due to better management strategy for reducing intra-interference~\cite{mao2018, joudeh2017multicast}. However, the HRSMA framework requires an additional SIC layer for decoding group common streams, resulting in additional cost and complexity at the user side. 
Second, users with the same or highly overlapped FoVs can be grouped together, resulting in reducing the sizes of FoV frames to be transmitter to the users. In the case that all the users in the same group, i.e., ``RSMA /wo clustering", the FoV frames transmitted to the users have to be large enough to maintain consistent experiences for the users.

\section{Conclusion}
\label{sec:conlcusion}
In this paper, we have proposed a novel joint communication and computation framework for 360-degree video streaming in virtual reality applications assisted by RSMA. In particular, in this framework, the users are clustered into different groups based on their interests, i.e., FoVs, and the HRSMA is utilized to transmit multicast streams to the users. As such, the users are served based on their specific interests, and the intra-interference among groups is efficiently managed.
We then introduced a highly-effective algorithm based on deep reinforcement learning to deal with the uncertainty and heterogeneity of the environment.
Extensive simulation results have demonstrated that our proposed solution can achieve millisecond-latency under the computing resource constraints of the transmitter and the dynamics of the wireless virtual reality environment as well as the users' demands. One of the potential research directions from this work is to incorporate users' FoV predicting algorithms, e.g., using deep learning, into the framework to reduce the sizes of FoV frames being transmitted over the wireless channels, thus further reducing the system's latency.

\end{document}